\documentclass[reprint, amsmath, amssymb, aps, pra, superscriptaddress]{revtex4-2}
\usepackage{xr-hyper}
\usepackage{style}

\newcommand{\Lbb}{\mathbb{L}}
\newcommand{\Nbb}{\mathbb{N}}

\newcommand{\Rbb}{\mathbb{R}}

\newcommand{\Zbb}{\mathbb{Z}}

\newcommand{\bbf}{\mathbf{b}}

\newcommand{\Gbf}{\mathbf{G}}

\newcommand{\tbf}{\mathbf{t}}

\newcommand{\wbf}{\mathbf{w}}

\newcommand{\Xbf}{\mathbf{X}}
\newcommand{\ybf}{\mathbf{y}}

\newcommand{\CalO}{{\mathcal{O}}}

\newcommand{\CalV}{{\mathcal{V}}}

\newcommand{\epf}{{\boldsymbol{e}}}
\newcommand{\fpf}{{\boldsymbol{f}}}
\newcommand{\Fpf}{{\boldsymbol{F}}}
\newcommand{\rpf}{{\boldsymbol{r}}}

\newcommand{\xpf}{{\boldsymbol{x}}}

\newcommand{\bsV}{\boldsymbol{V}}
\newcommand{\bsW}{\boldsymbol{W}}

\newcommand{\sigpf}{\pmb{\sigma}}
\newcommand{\phipf}{\pmb{\phi}}
\newcommand{\Phipf}{\pmb{\Phi}}

\newcommand{\ind}[1]{\mathbbm{1}_{#1}}

\renewcommand{\tilde}[1]{\widetilde{#1}}

\newcommand{\lan}{\left\langle}
\newcommand{\ran}{\right\rangle}

\newcommand{\wstar}{\wbf^\star}
\newcommand{\what}{{\widehat{\wbf}}}

\begin{document}

\preprint{APS/123-QED}

\title{Physics-guided weak-form discovery of reduced-order models for trapped ultracold hydrodynamics}

\author{Reuben R. W. Wang}
 \affiliation{JILA, NIST, and Department of Physics, University of Colorado, Boulder, Colorado 80309, USA}
 \email{reuben.wang@colorado.edu}
\author{Daniel Messenger}%
\affiliation{Department of Applied Mathematics, University of Colorado Boulder.}%
 \email{daniel.messenger@colorado.edu}

\date{\today}

\begin{abstract}
We study the relaxation of a highly collisional, ultracold but nondegenerate gas of polar molecules. 
Confined within a harmonic trap, the gas is subject to fluid-gaseous coupled dynamics that lead to a breakdown of first-order hydrodynamics.  
An attempt to treat these higher-order hydrodynamic effects was previously made with a Gaussian ansatz and coarse-graining model parameter [R. R. W. Wang \& J. L. Bohn, \href{https://journals.aps.org/pra/abstract/10.1103/PhysRevA.108.013322}{Phys. Rev. A 108, 013322 (2023)}], leading to an approximate set of equations for a few collective observables accessible to experiments.
Here we present substantially improved reduced-order models for these same observables, admissible beyond previous parameter regimes, discovered directly from particle simulations using the WSINDy algorithm (Weak-form Sparse Identification of Nonlinear Dynamics). The interpretable nature of the learning algorithm enables estimation of previously unknown physical quantities and 
discovery of model terms with candidate physical mechanisms,
revealing new physics in mixed collisional regimes. Our approach constitutes a general framework for data-driven model identification leveraging known physics.
\end{abstract}

\maketitle

Recent experiments have demonstrated the capacity to collisionally shield ultracold polar molecules against inelastic loss \cite{Matsuda20_Sci, Valtolina20_Nat, Anderegg21_Sci, Schindewolf22_Nat, Lin23_PRX, Bigagli23_NatPhys}, either with the application of static \cite{Quemener16_PRA, Gonzalez17_PRA, Lassabliere22_PRA} or microwave electric fields \cite{Avdeenkov12_PRA, Lassabliere18_PRL, Karman18_PRL, Karman20_PRA}.
This achievement has allowed direct evaporative cooling of highly polar molecular gases in three-dimensions with collisional thermalization to quantum degenerate temperatures \cite{Li21_Nat, Schindewolf22_Nat, Bigagli23_arxiv}, ushering in a new era for quantum gases. 
Unfortunately, the blessing of large dipole moments for exploration of long-ranged physics also presents a curse for evaporative cooling. Larger dipoles result in higher collision rates, bringing the gas into a hydrodynamic regime where re-equilibration is significantly delayed compared to its dilute counterpart due to fluid excitations \cite{Ma03_JPB, Wang23_PRA2}. 
Access to simple models of these collective excitations could, therefore, aid in the design of more efficient evaporation protocols, motivating their discovery. 

Deriving models of collective dynamics in many-body systems has a long history, dating back at least as far as Boltzmann's analysis of thermal gases \cite{Boltzmann1871_KHS}. 
In hard-boundary containers, the evolution of strongly collisional gases is well predicted using continuum mechanics, whereby coupled equations for the density, velocity, and temperature fields constitute an effective continuum model \cite{Fetter03_Dover}.
However, soft-boundary conditions introduce the complication of mixed collisional regimes \cite{Bruun07_PRA, Schafer12_Springer, Schafer14_PRA}, whereby hydrodynamic (collisionally thick) and dilute (collisionally thin) regions now undergo coupled dynamics.
Nevertheless, symmetries of the system, even if only approximate, can be exploited to obtain a reduced-order model (ROM) that governs a small number of relevant collective observables \cite{messenger2023coarse}. Traditionally, these ROMs would be obtained with a judicious choice of an ansatz that appropriately preserves the system's symmetries \cite{Wang23_PRA2}, and so are restricted by the assumptions made with the ansatz. Modern tools from machine learning (e.g.\ neural networks) can yield highly expressible models \cite{hernandez2021deep}, but are computationally prohibitive and difficult to interpret due to latent variable representations.

In this Letter, we implement a physics-informed machine-assisted approach for discovering ROMs of trapped, ultracold dipolar gases in the hydrodynamic regime.
Compared to an ansatz-based ROM previously derived in Ref.~\cite{Wang23_PRA2}, the ROMs learned here for the same experiments show a drastic reduction in errors against the ground-truth dynamical data \footnote{The term ground-truth data refers to data taken from a real (or at least realistic) generating source, serving as a benchmark for training the machine learning model. }, while remaining physically interpretable. 
The latter also shows successful dynamical predictions in regimes outside of those accessible to the former, showcasing the robustness of our method for more comprehensive parameter studies. 
Although applied to numerical simulation data, we expect our data-driven learning technique will naturally extend to real data from ultracold trapped gas experiments, where accurate and repeatable thermometry measurements can be taken with the high degree of control possible in ultracold systems \cite{Li21_Nat, Patscheider22_PRA, Schindewolf22_Nat}.

Our primary tool for ROM identification is weak-form equation learning, whereby human-readable governing equations are identified in weak form to describe a given set of observables. The interpretable nature of equation learning has enabled physics discovery from experimental data in a variety of disciplines, including material science, nano-engineering, and robotics \cite{naik2022discovering,johnston2022equation,kaheman2019learning}. In contrast to expensive forward solver-based approaches (e.g.\ nonlinear least squares \cite{banks2012estimation}), or black-box optimization routines (e.g.\ deep learning \cite{LeCun15_Nat}), equation learning bypasses computational burdens by discretizing equations directly using available data \cite{Brunton16_PNAS}. The weak form is then employed to distill reduced laws from data exhibiting fluctuations around a suitable reduced description \cite{messenger2020weak,messenger2021weak}. In practice, access to all dynamical variables of interest (e.g.\ density, temperature, and flow fields) is challenging for experiments, thus the state space for learning must be limited to realistic observables. In the current study, our technique identifies a closed set of equations for a low-dimensional set of observed variables (the root-mean-squared spatial widths of the gas) that reproduces the data derived from high-dimensional simulations with excellent quantitative accuracy, improving on previous work \cite{Wang23_PRA2}, despite the significant reduction in dimensionality from the full state space.

{\it Trapped Hydrodynamics.}
We focus on cross-dimensional rethermalization experiments \cite{Monroe93_PRL, Wang21_PRA}, designed to probe the re-equilibration rate of a gas and, consequently, the rate at which forced evaporation of molecules should occur. 
These experiments are performed at temperatures above quantum degeneracy where evaporation typically commences, and consists of a near-instantaneous initial excitation of the gas along one direction before allowing it to relax.  
When nondegenerate, molecular motion is well described by classical trajectories so that a collisionally-thick molecular gas, which rapidly thermalizes the local momentum distribution, is likened to a classical thermoviscous fluid \cite{Griffin97_PRL, Nikuni98_JLTP, Kavoulakis98_PRA}.    
Gas dynamics is then best described by the equations of fluid mechanics \cite{Fetter03_Dover} (see SI Sec. 1), incorporating the transport tensors of thermal conductivity $\kappa_{i j}$ and viscosity $\mu_{i j k \ell}$ \cite{Wang22_PRA, Wang22_PRA2, Wang23_PRA} arising from quantum dipolar collisions \cite{Bohn09_NJP}. The tensor subscripts denote coordinate indices. 

Unfortunately, the fluid description can get muddied in optical dipole traps \cite{Grimm00_AAMOP} that confine these molecular samples, as they give rise to soft boundary conditions. In particular, deep optical dipole traps that confine molecules of mass $m$ are well approximated by a cylindrically symmetric harmonic potential 
$V(\boldsymbol{r}) = m \omega_{\perp}^2 ( x^2 + y^2 + \lambda z^2 )/2$, where $\omega_{\perp}$ is the radial trapping frequency and $\lambda = (\omega_z / \omega_{\perp})^2$ is the trap aspect ratio, 
indeed causing the gas density to decay away from the trap center. Sufficiently far from the trap center, we expect that the gas is mostly dilute, leading to the presence of a fluid-gaseous interstitial region.    
Ref.~\cite{Wang23_PRA2} attempts a reductionist approach to this issue, by employing an empirically determined stationary coarse-graining parameter, the hydrodynamic volume ${\cal V}_{\rm hy}$, to subsume all multiphase fluid dynamics \footnote{ A similar treatment has been attempted by Ref.\cite{Kavoulakis98_PRA} for isotropic scatterers. }. This approximation allowed the derivation of a reduced-order model for cross-dimensional rethermalization dynamics with a Gaussian ansatz
$\rho(\boldsymbol{r}, t) 
=
m N
\prod_{i=1}^3
\exp( -\frac{ r_i^2 }{ 2 \sigma_i^2(t) } )
/ \sqrt{ 2 \pi \sigma_i^2(t) }$ to describe the density profile.
As a result, the trapped gas was found to have its anisotropic Gaussian widths $\sigma_i(t)$ governed approximately by the ordinary differential equations (ODEs)
\begin{align} \label{eq:widths_ODE_Tindependent}
    {\rm E}_i(\boldsymbol{\sigma}, \dot{\boldsymbol{\sigma}}, \ddot{\boldsymbol{\sigma}})
    -
    \frac{ 2 k_B T_0 }{ m \sigma_i } 
    =
    - \frac{ 2 }{ 5 }
    \frac{ {\cal V}_{\rm hy} }{ N m } 
    \sum_{j} 
    \frac{ \mu_{i i j j}( T ) }{ \sigma_i }
    \frac{ \dot{\sigma}_{j} }{ \sigma_{j} },
\end{align}
where ${\rm E}_i(\boldsymbol{\sigma}, \dot{\boldsymbol{\sigma}}, \ddot{\boldsymbol{\sigma}}) := \ddot{\sigma}_i + \omega_i^2 \sigma_i + \frac{ 1 }{ 3 \sigma_i } \sum_j ( \omega_j^2 \sigma_j^2 + \dot{\sigma}_j^2 )$.  
Above, $T_0$ is the equilibrium temperature, $N$ the number of molecules and $k_B$ is Boltzmann's constant. 

The Supplementary Information (SI) provides further details on deriving the model above, which asserts the ideal gas law equation of state, an approximately spatially uniform viscosity over the sample, and a particular flow velocity field given by $U_i(\boldsymbol{r}, t) = [ \dot{\sigma}_i(t) / \sigma_i(t) ] r_i$. 

While Eq.~\eqref{eq:widths_ODE_Tindependent} provides a useful approximation, the main contribution of this work is to demonstrate the utility of weak-form equation learning in correcting model componentssuch as $\CalV_\text{\rm hy}$, and generalizing Eq.~\eqref{eq:widths_ODE_Tindependent} by discovering new model terms. 
We expect the newly learned model terms to arise predominantly from second-order hydrodynamics \cite{Bluhm15_PRA, Lewis17_IOP}, where the transport tensors inherit a time-dependence following a relaxation law with timescale $\tau_{\mu}$, proportional to the viscosity and inverse density \cite{Schafer14_PRA}. This latter dependence results in anisotropic relaxation timescales of the transport tensors due to dipolar collisions, complicating previous treatments. 
Our machine-assisted, physics-informed approach is able to overcome these complications, enabling discovery of effective ROMs for the trapped gas dynamics.

{\it Weak-form equation learning}. 
We perform model discovery with
weak-form equation learning (WFEL), which attempts to learn a representation of a differential equation in a suitable {\it weak formulation} using discretely sampled solution data. For an ODE  $\dot{\xpf}(t) = \fpf(\xpf(t))$ in $d$ dimensions with initial data $\xpf(0) = \xpf_0\in \Rbb^d$, its weak formulation is obtained by integrating against a smooth, compactly-supported {\it test function} $\phipf:(0,\infty)\to \Rbb^d$ and using integration by parts to get 
\begin{equation}\label{weakform}
-\int_0^\infty \dot{\phipf}(t)\cdot \xpf(t)\,dt = \int_0^\infty\phipf(t)\cdot \fpf(\xpf(t))\,dt.
\end{equation}
When equation \eqref{weakform} holds for all continuously differentiable $\phipf$, we say that $\xpf$ is a {\it weak solution} to the ODE. The objective of WFEL is to identify the vector field $\fpf$ governing the dynamics of $\xpf$ by leveraging Eq.~\eqref{weakform}. 

WSINDy extends the SINDy framework (Sparse Identification of Nonlinear Dynamics) \cite{Brunton16_PNAS} to accomplish this task, whereby $\fpf$ is assumed to satisfy $\fpf_i(\xpf) = \sum_{j=1}^J\wstar_{ji} f_j(\xpf) := \Lbb(\xpf) \wstar_i$ for a sparse {\it weight matrix} $\wbf^\star\in \Rbb^{J\times d}$ and a library of {\it trial functions} $\Lbb = \{f_j:\Rbb^d \to \Rbb\}_{j=1}^J$. We use the notation $\Lbb(\xpf)$ to refer to the {\it quasi-matrix} with columns given by the functions $f_j(\xpf(\cdot)):\Rbb \to \Rbb$. A weak formulation for inference is defined by specifying a finite set of test functions $\Phipf = \{\phi_1,\dots,\phi_K\}$ and computing a linear system $(\Gbf,\bbf)$ according to \eqref{weakform}, with 
\begin{equation} \label{eq:weakform_disc}
    \Gbf_{kj} = \int_0^\infty \phi_k(t) f_j(\xpf(t))\,dt, \:\:\:
    \bbf_{ki} = -\int_0^\infty \dot{\phi}_k(t)\xpf_i(t)\,dt.
\end{equation} 
As defined above, $\Gbf \wstar = \bbf$ holds exactly on weak solutions $\xpf$. In practice, $(\Gbf,\bbf)$ is computed from corrupted data $\ybf = \xpf(\tbf)+\epf\in\Rbb^{M\times d}$ sampled on a timegrid $\tbf = (t_1,\dots,t_M)$, where $\xpf$ is a (weak) solution to $\dot{\xpf}(t) = \Fpf(\xpf(t))$. The corruptions $\epf$ may encapsulate measurement noise or model error. In the current setting, $\epf$ is associated with Monte Carlo sampling error of physical quantities with numerical methods discussed later.  
The integrals in \eqref{eq:weakform_disc} are computed from observations $\ybf$ using the trapezoidal rule, a highly accurate scheme when $\epf=0$ and $\phi_k$ decays smoothly to the edge of its support \cite{messenger2020weak}. An approximate solution $\what\approx \wstar$ is then found by solving $\Gbf\what\approx \bbf$ using sparse regression \cite{Brunton16_PNAS}, as outlined in SI Sec.~2.

\vspace{-0.03cm}{\it Learning with physical knowledge.} 
We systematically incorporated knowledge of the underlying physics into the optimization problem for $\what$, constituting three learning strategies: 
{\it na\"ive discovery} (ND), {\it ansatz verification} (AV), and {\it physics-informed} (PI) learning. These strategies enforce different degrees of physical knowledge to a) inform the library $\Lbb$ and b) constrain some coefficients in $\what$ using the ansatz \eqref{eq:widths_ODE_Tindependent}. See Table \ref{tab:libraries} for choices of $\Lbb$.
Resulting WSINDy models will be referred to as WSINDy-ND, WSINDy-AV, and WSINDy-PI, respectively. 

In the ND approach, a standard monomial library is used for $\Lbb$ and $\what$ is left unconstrained. For the AV and PI approaches, $\Lbb$ is motivated from known approximations used to derive \eqref{eq:widths_ODE_Tindependent}, while several trial functions $f_j$ are removed from the learning process, with their respective coefficients $\what_{ji}$ restricted to values derived in \eqref{eq:widths_ODE_Tindependent}. Specifically, let $\wbf^{\text{Ans}}$ be the weight matrix of coefficients from the ansatz model \eqref{eq:widths_ODE_Tindependent}, and let $S_i^0\subset\{1,\dots,J\}:=[J]$ be the index set for terms removed from the learning process in the $i$th equation. We then let $\Gbf^{(i)} = \Gbf_{(S_i^0)^c}$, $\bbf^{(i)} = \bbf_i -  \Gbf_{S_i^0}\wbf^{\text{Ans}}_{i,S_i^0}$
where $\Gbf_{(S_i^0)^c}$ is the restriction of $\Gbf$ to the column indices $(S_i^0)^c = [J]\setminus S_i^0$, and $\wbf^{\text{Ans}}_{i,S_i^0}$ is the restriction of the $i$th column of $\wbf^{\text{Ans}}$ to the indices $S_i^0$. Each $i\in [d]$, is then solved with sparse regression,
$\min_{\wbf^{(i)}\in\Rbb^{J^{(i)}}}\| \Gbf^{(i)}\wbf^{(i)} -\bbf^{(i)} \|_2^2 +\nu^2\|\wbf^{(i)}\|_0$ (see SI).

\onecolumngrid

\begin{table}[h]
    \caption{\label{tab:libraries}  
    Libraries utilized in each learning strategy. For $\Lbb^N$ we use a standard monomial library with additional damping terms that can be computed without explicit computation of $\dot{\sigma}_i$ using the weak form. $\Lbb^{AV}$ contains the terms in the ansatz \eqref{eq:widths_ODE_Tindependent} not appearing in the LHS of \eqref{eq:ansatz_AVlib}. For $\Lbb^{f}$ we used $p_{\max}=3$ to balance expressivity and accuracy of the learned models. 
    }
    \begin{ruledtabular}
    \begin{tabular}{lc}
    \textrm{ Learning strategy } &
    \textrm{ Library } \\
    \colrule 
    {Na\"ive Discovery} &     
    $\displaystyle\Lbb^N := \left\{(\sigma^{p_1}_1\cdots\sigma^{p_d}_d),\,\frac{d}{dt}(\sigma^{p_1}_1\cdots\sigma^{p_d}_d)\ :\  1\leq\sum_i p_i\leq 5, p_i\in\Nbb\right\}$
    \\[1.5em]
    {Ansatz Verification} & $\displaystyle\Lbb_i^{AV}:= \left\{\frac{1}{\sigma_i(t)}, \left(\frac{\dot{\sigma_j}(t)}{\sigma_i(t)\sigma_j(t)}\right)_{j=1}^d\right\}$ 
    \\[1.5em] 
    {Physics-Informed} & $\displaystyle\Lbb_i^\mu := \left(\frac{\dot{\sigma_j}(t)}{\sigma_j(t)\sigma_k(t)},\frac{\dot{\sigma_i}(t)\sigma_j(t)}{\sigma_i(t)},\frac{\dot{\sigma_i}(t)\dot{\sigma_j}^2(t)}{\sigma_k(t)\sigma_j(t)}\right)_{j,k=1}^d$,
    \quad $\Lbb^{f} := \left\{(\sigma^{p_1}_1\cdots\sigma^{p_d}_d)\ :\ \sum_i\left\vert p_i\right\vert \leq p_{\max}, p_i\in\Zbb\right\}$
    \end{tabular}
    \end{ruledtabular}
\end{table}

\twocolumngrid

{\it Monte Carlo experiments.} As a proxy for real experimental data, we perform simulations of the gas dynamics by numerical integration of the Boltzmann equation \cite{Reif09_Waveland}. 
Treating the particles as discrete points in classical phase-space, they are first sampled from the initial out-of-equilibrium distribution, then evolved in discrete time via the St\"ormer-Verlet symplectic integrator \cite{Hairer03_AN}. 
Two-body collisions between particles are sampled using the direct simulation Monte Carlo (DSMC) method \cite{Bird70_PF, Sykes15_PRA, Wang23_PRA2}, which exploits the locality of interactions for computational efficiency. 
In our implementation, the simulation volume is first adaptively partitioned into discrete grid cells of local density-dependent volumes, into which the simulated molecules are binned based on their positions. Collisions are then sampled within each grid cell based on the scattering cross section between ultracold dipoles \cite{Bohn14_PRA}, leading to updates in the molecular-pair momenta. For all our simulations, we take the molecular dipoles to be polarized along $\hat{\boldsymbol{z}}$, enforcing cylindrical symmetry of the system in tandem with the trap $V(\boldsymbol{r})$.
We leverage previous findings from Ref.~\cite{Wang23_PRA2} and restrict learning to be done on $\sigma_i^2(t) = \sum_k (r_i^{(k)})^2 / N$ from the simulation particle ensemble, summed over simulated particles indexed by $k$.  
These observables mimic experimental data that would be extracted from in-situ absorption imaging of the molecular gas cloud \cite{Wang10_PRA}, permitting a natural extension to learning on real experimental data.

Taking the DSMC computed time traces as ground-truth data \footnote{ The simulation parameters are those used in Ref.~\cite{Wang23_PRA2}. }, we now examine the three approaches for reduced-order model learning in more detail. 
Initial out-of-equilibrium excitations are always along $z$, further enforcing the cylindrical symmetry about $z$. The time traces for $\sigma_{x}$ and $\sigma_y$ can be subsequently averaged to obtain $\sigma_\perp := (\sigma_x^2+\sigma_y^2)^{1/2}$. We present 2nd-order (in time) models learned from $(\sigma_\perp,\sigma_z)$, so that $d=2$ in what follows. Learning is done on adimensional variables, rescaled by natural units (see SI Sec. 2.1).

To assess the performance of each model, we measure the relative standard deviation $\Delta_2(\widehat{\sigma}_i) = \text{std}(\widehat{\sigma}_i-\sigma_i)/\text{std}(\sigma_i)$, where $\widehat{\sigma}_i$ represents the time trace obtained from ROM solutions (ansatz or WSINDy model), and ${\sigma}_i$ represents the corresponding DSMC time trace. We find $\Delta_2$ to be more informative than the standard relative $\ell_2$ error, which is misleadingly small due to the small amplitude nature of fluctuations in the widths $\sigma_i$.

{\it Na\"ive Discovery (ND).} 
As a benchmark against the physics-informed approaches to follow, we first attempt to learn the dynamics of the observed widths $\sigpf = (\sigma_i)_{i=1}^d$ with no knowledge of the underlying physics, using
\begin{equation}\label{eq:ND}
    \ddot{\sigma}_i = \Lbb_i(\sigpf) \wbf_i^{ND}
\end{equation}
with libraries $\Lbb_i=\Lbb^N$ of generalized forcing and damping terms represented in monomial bases (see Table \ref{tab:libraries}). The resulting WSINDy-ND models confirm that the widths can be described as a 2nd-order nonlinear damped oscillator, with forward simulation errors on par with the ansatz (Fig.\ \ref{fig:PIres}). However, like the ansatz, performance is highly $\lambda$-dependent, with underresolved $\sigma_z$ ($\sigma_\perp$) dynamics for small (large) $\lambda$. In addition, the library $\Lbb^N$, while explicit, is less amenable to interpretation.

{\it Ansatz Verification (AV).} 
Next, we perform a physics-directed learning strategy that directly assesses the validity of the ansatz in Eq.~\eqref{eq:widths_ODE_Tindependent}. In this approach, WSINDy is employed to estimate ansatz parameters in terms most likely to have been underresolved, namely the equation of state (EOS) and viscous terms.  
The ideal gas law EOS produces the term $-\frac{2k_BT_0}{m\sigma_i(t)}$ in each equation for $\sigma_i$. Meanwhile, viscous forcing terms in \eqref{eq:widths_ODE_Tindependent} are derived under the combined assumptions that a) the spatial variation of the temperature field in $\mu_{0}(T)$ is negligible, and b) there exists a hydrodynamic volume $\CalV_\text{hy}$ within which dynamical effects of viscosity can be ignored.  
To this end, we restrict the WSINDy library to the terms appearing in the ansatz \eqref{eq:widths_ODE_Tindependent}, and we enforce the exact coefficients for terms associated with quantities that follow immediately from the fluctuating Gaussian ansatz to $\rho(\boldsymbol{r}, t)$ with corresponding flow field $U_i = [ \dot{\sigma}_i / \sigma_i ] r_i$, and known properties of $V(\boldsymbol{r})$. That is, we seek a representation  
\begin{align} \label{eq:ansatz_AVlib}
    {\rm E}_i(\boldsymbol{\sigma}, \dot{\boldsymbol{\sigma}}, \ddot{\boldsymbol{\sigma}})
    =
    \Lbb_i(\sigpf)\wbf_i^{AV}, 
\end{align}
with libraries $\Lbb_i=\Lbb_i^{AV}$ given in Table \ref{tab:libraries} [see also \eqref{eq:widths_ODE_Tindependent}].

\begin{figure}[ht]
    \centering 
    \includegraphics[clip,trim={0 5 0 0},width=\columnwidth]{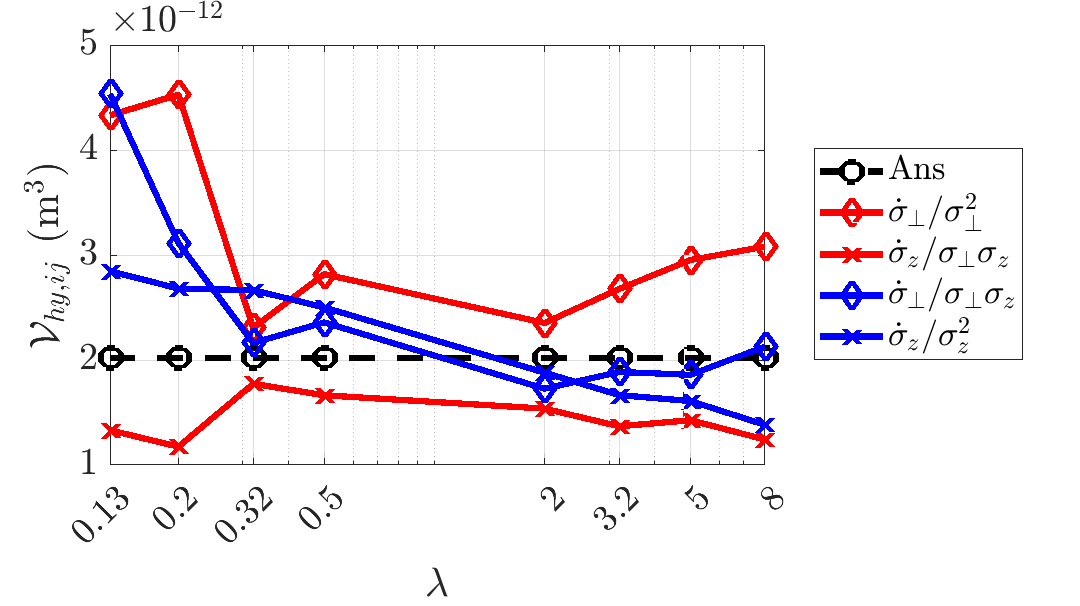} 
    \vspace{-15pt}
    \caption{ Learned ${\cal V}_{{\rm hy}, ij}$ (Eq.\ \eqref{eq:Vhydij}) using the WSINDy-AV models. The empirical functional form for ${\cal V}_{\rm hy}$ \cite{Wang23_PRA2} is independent of $\lambda$ and given by the black dashed line  (SI Eq. 1.11). 
    Red and blue curves represent values extracted from the learned equations for $\ddot{\sigma}_\perp$ and $\ddot{\sigma}_z$, showing anisotropies in the hydrodynamic volume at different trap ratios $\lambda$. Modest gains in accuracy indicate that such anisotropy is favorable (Fig.~\ref{fig:PIres}).}
    \label{fig:Vhyd}
\end{figure}

\begin{figure}[ht]
    \centering 
    \includegraphics[clip,trim={15 5 20 10},width=\columnwidth]{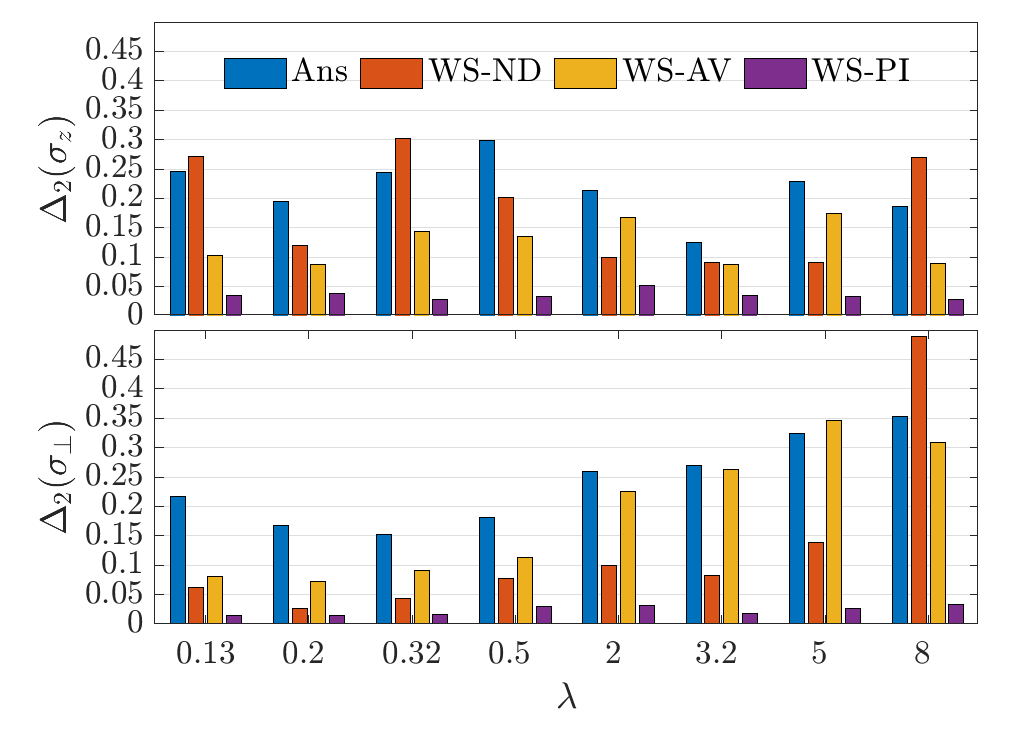}
    \vspace{-25pt}
    \caption{Fidelity of the ansatz (blue), WSINDy-ND (red), WSINDy-AV (yellow), and WSINDy-PI (purple) models with respect to DSMC data. WSINDy-PI captures the dynamics to within $5\%$ across the spectrum of trap aspect ratios $\lambda$, reducing errors by an order of magnitude from the ansatz.}
    \label{fig:PIres}
\end{figure}

\begin{figure}[ht]
    \includegraphics[trim={25 3 35 5},clip,width=\columnwidth]{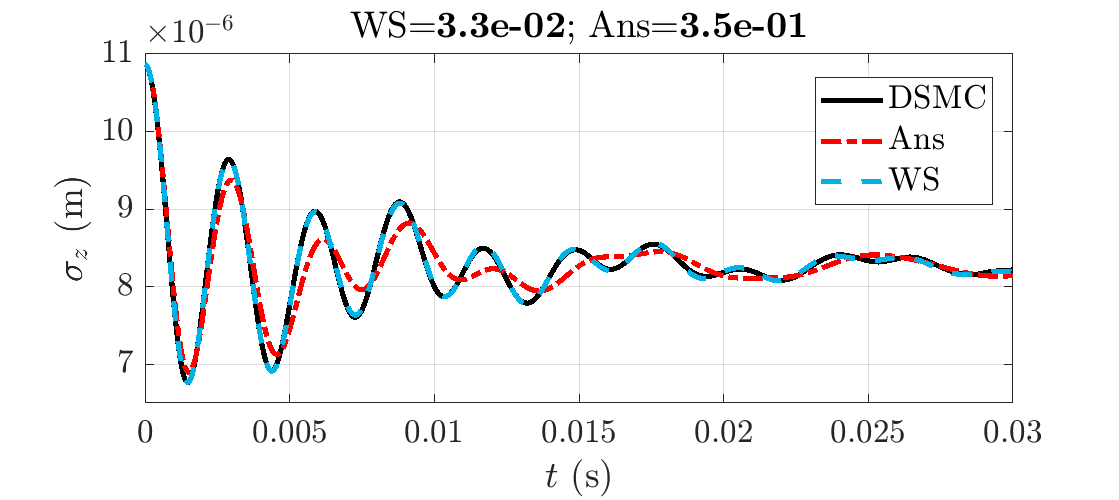}
    \vspace{-20pt}
    \caption{ 
    Comparison between DSMC, Ansatz, and WSINDy-PI for $\lambda=8$ (pancake-shaped trap). 
    The relative errors of the WSINDy (WS) and Ansatz (Ans) time traces with respect to DSMC results are listed in the titles.}
    \label{fig:timetrace_lambda10}
\end{figure}

The WSINDy-AV models for each $\lambda$ validate the ideal gas law to 4 significant digits, comparable to the $1/\sqrt{N}$ Monte Carlo accuracy, and estimate the hydrodynamic volume ${\cal V}_{\rm hy}$ to be of order $\CalO(10^{-12} {\rm m}^3)$, in agreement with previous empirical findings \cite{Wang23_PRA2} (see Eq. 1.11 and Table 1 
of the SI). We moreover identify a distinct anisotropy in ${\cal V}_{\rm hy}$ dependent on $\lambda$. 
By relaxing the assumption of a scalar $\CalV_\text{hy}$, 
we extract its anisotropic generalization by replacing the right-hand side of \eqref{eq:widths_ODE_Tindependent} with  
\begin{equation} \label{eq:Vhydij}
    \CalV_{\text{hy}, ij} 
    =
    - \frac{5Nm}{2\mu_{iijj}}c_{ij},
\end{equation} 
where $c_{ij}$ are coefficients of $\dot{\sigma}_j (\sigma_i \sigma_j)^{-1}$ in Eq.~(\ref{eq:widths_ODE_Tindependent}) learned by WSINDy.
Fig.~\ref{fig:Vhyd} visualizes the trend in $\CalV_{\text{hy}, ij}$ vs $\lambda$ for each viscosity coefficient. Principally, the red and blue curves, together with the modest gains in accuracy depicted in Fig.~\ref{fig:PIres}, indicate that $\CalV_{\text{hy}, ij}$ better captures the true dynamics. 
However, such adjustments do not correct for the drifting oscillatory phase
observed and amplitude attenuation of $\sigma_i(t)$, most present at larger $\lambda$ as seen in Fig.~\ref{fig:timetrace_lambda10}. 
Our final physics-informed approach corrects these issues by expanding the viscous force library, arriving at ROMs viable across the spectrum of $\lambda$ values.

{\it Physics-Informed Discovery (PI).} 
With our final strategy, we aim to find expressions for higher-order hydrodynamic 
effects omitted from the ansatz-based model that hold across $\lambda$ and correct the previously noted phase drift and amplitude attenuation. We expect that primary sources of corrections to Eq.~\eqref{eq:widths_ODE_Tindependent} arise from the significant dilute fraction interfering with hydrodynamic excitations. 
From the AV study, we may conclude that the leading order EOS terms $-\frac{2 k_B T_0}{m \sigma_i(t)}$ are correctly modeled by the ansatz. We therefore seek a representation for the entire left-hand side of Eq.~\eqref{eq:widths_ODE_Tindependent} that includes corrections from higher-order hydrodynamics:
\begin{align} \label{eq:ansatz_lib} 
    \hspace{-2.2mm}{\rm E}_i(\boldsymbol{\sigma}, \dot{\boldsymbol{\sigma}}, \ddot{\boldsymbol{\sigma}})
    -
    \frac{2k_BT_0}{m\sigma_i(t)} 
    &= 
    \Lbb^{f}(\sigpf)\wbf_i^{PI,f}+\Lbb_i^{\mu}(\sigpf)\wbf_i^{PI,\mu}
\end{align}
where the augmented viscous libraries $\Lbb_i^{\mu}$ and forcing library $\Lbb^{f}$ are defined in Table \ref{tab:libraries}. 
Terms in $\Lbb^{f}$ are motivated by the EOS and the restorative forces imparted by the harmonic trap, both of which involve only $\pmb{\sigma}$. Terms in $\Lbb_i^{\mu}$ depend on both $\pmb{\sigma}$ and $\dot{\pmb{\sigma}}$ that vanish as $\dot{\pmb{\sigma}}\to 0$, and are motivated by considering a first-order Taylor expansion of ${\cal V}_{\rm hy}$ around the equilibrium temperature ${\cal V}_{\rm hy}( \langle T \rangle )
= 
{\cal V}_{\rm hy}( T_0 )
+
\left. 
{ \partial {\cal V}_{\rm hy} / \partial \langle T \rangle }
\right|_{\langle T \rangle = T_0}
( \langle T \rangle - T_0 )$. 
With $\langle T \rangle = 2 T_0 - \frac{ m }{ 3 k_B } \sum_k \left[ \omega_k^2 \sigma_k^2(t) + \dot{\sigma}_k^2(t) \right]$, following from the ansatz for $\rho(\boldsymbol{r}, t)$ and $\boldsymbol{U}(t)$, the libraries $\Lbb^\mu_i$ are naturally suited to capture corrections to ${\cal V}_{\rm hy}$. 
Here $\lan \cdot\ran$ denotes an averaging over the molecular distribution.

Fig.~\ref{fig:PIres} depicts significant improvements of the WSINDy-PI models (purple bars) compared to the ansatz model (blue bars), with contributions from $\Lbb^f$ and $\Lbb^\mu$ capturing the DSMC data to within (and often far below) 5\% across all $\lambda$ values, reducing errors by an order of magnitude. The resulting WSINDy-PI models preserve equilibria to within the tolerance of the DSMC data (i.e.\ within $0.1\%$, comparable to $1/\sqrt{N}$), due to 
corrections from $\Lbb^f$ effectively canceling as $t\to\infty$, permitting each $\sigma_i$ to reach their thermal equilibrium value as necessitated by Boltzmann's H-theorem \cite{Reif09_Waveland}.

\begin{figure}[ht]
    \includegraphics[trim={25 3 35 5},clip,width=\columnwidth]{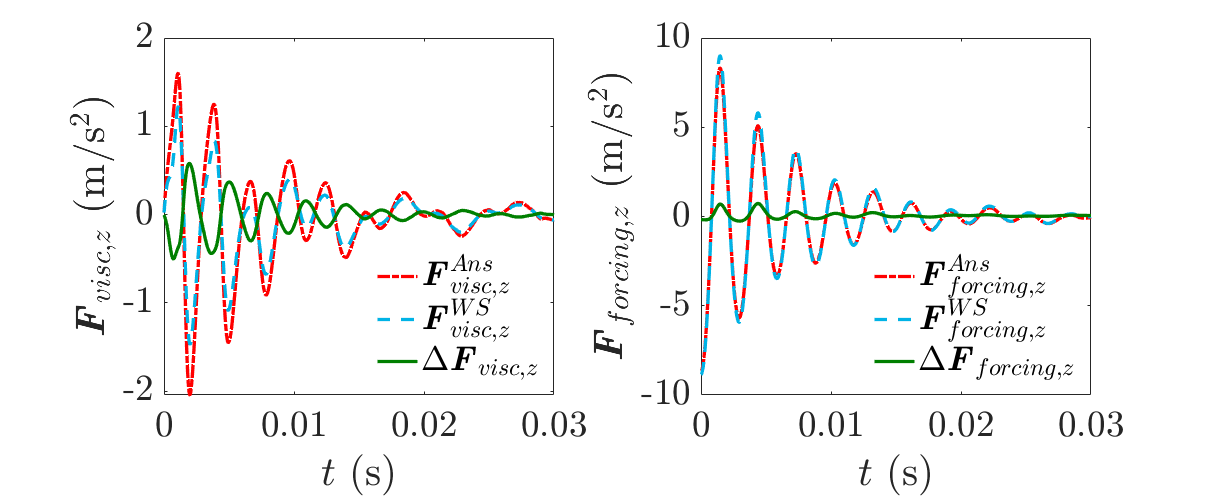}
    \vspace{-20pt}
    \caption{ 
    Comparison of the specific forcing time traces along $\sigma_z$ between the Ansatz (dot-dashed red curve), and WSINDy-PI (dashed blue curve) for $\lambda=8$ (pancake-shaped trap), for both the viscous (left) and forcing (right) components. 
    The difference between the WSINDy and Ansatz forcing terms $\Delta\boldsymbol{F}_z = \boldsymbol{F}^{WS}_z - \boldsymbol{F}^{Ans}_z$ is plotted in solid green. }
    \label{fig:timetrace_forcings}
\end{figure}

Improvements imparted in the WSINDy-PI models can be understood by isolating terms that appear in the ansatz model and splitting the remaining terms into ``forcing'' and ``viscous'' components that correct the specific (per unit mass) forcing terms from Ref.~\cite{Wang23_PRA2}: 
\begin{align}
    \ddot{\pmb{\sigma}} 
    = 
    \pmb{F}_{\text{Ans}}(\pmb{\sigma},\dot{\pmb{\sigma}}) 
    +
    \pmb{F}_{\text{forcing}}(\pmb{\sigma})
    +
    \pmb{F}_{\text{visc}}(\pmb{\sigma},\dot{\pmb{\sigma}}).
\end{align}
The $\sigma_z$ components of $\pmb{F}_{\text{forcing}}$ and $\pmb{F}_{\text{visc}}$ are visualized for $\lambda=8$ (pancake-trap) in Fig.\ \ref{fig:timetrace_forcings}, where we observe that the additional learned viscous force $\Delta \pmb{F}_{\text{visc}} = \pmb{F}^{WS}_{\text{visc}} - \pmb{F}^{Ans}_{\text{visc}}$ (left plot) opposes the ansatz model, resulting in a weakened viscous damping, while $\Delta \pmb{F}_{\text{forcing}} = \pmb{F}^{WS}_{\text{forcing}} - \pmb{F}^{Ans}_{\text{forcing}}$ provides additional restorative force. These corrective specific forces act to dynamically modify the phase and amplitude of the observed weltering oscillations, visualized in Fig.~\ref{fig:timetrace_forcings}, with $\pmb{F}_{\text{visc}}$ acting to correct oscillation amplitudes and $\pmb{F}_{\text{forcing}}$ correcting oscillation frequencies (see Fig.\ \ref{fig:timetrace_lambda10}).

The WSINDy-PI improvements can be readily interpreted by considering the modifications they impart to the linearized equations of motion
\begin{align} \label{eq:linearized_widths_EOM}
    \ddot{\delta\sigma}_i(t)
    +
    2 \sum_j 
    \Gamma^{\rm PI}_{i j}
    \dot{\delta\sigma}_{j}(t)
    +
    \sum_j
    O^{\rm PI}_{i j}
    \delta\sigma_j(t)
    &\approx
    0,
\end{align}
that model small perturbations $\delta\sigma_i(t) = \sigma_i(t) - \sigma_i(\infty)$, from equilibrium.
Learned terms modify the squared-frequency and damping matrices according to 
\begin{subequations}
\begin{align}
    O^{\rm PI}_{i j}
    &=
    \left[
    2 \omega_i^2 \delta_{i,j} 
    +
    \frac{ 2 }{ 3 } 
    \omega_i \omega_j
    \right]  
    \frac{\partial (\Fpf_\text{forcing})_i}{\partial \sigma_j} (\pmb{\sigma}(\infty)),  
    \\
    \Gamma^{\rm PI}_{ i j }
    &= 
    \left[
    \frac{ \mu_0 {\cal V}_{\rm hy} }{ 5 N k_B T_0 }
    \omega_i
    M_{i j} 
    \omega_{j}
    \right] 
    +
    \frac{\partial (\Fpf_\text{visc})_i}{\partial \dot{\sigma}_j} (\pmb{\sigma}(\infty),\pmb{0}),
\end{align}
\end{subequations} 
identified as explicit corrections to the mode oscillation frequencies and their viscous damping rates, with those originally from Eq.~(\ref{eq:widths_ODE_Tindependent}) in square brackets. For example, at $\lambda=8$, corrections to the squared-frequencies acting on $\sigma_z$,  $O_{21}^\text{PI}$ and  $O_{22}^\text{PI}$, are approximately 
$4 \pi^2 1.48\times 10^2$ Hz$^2$ and $4 \pi^2 7.78 \times 10^3$ Hz$^2$, 
representing 1.6\% and 9.7\% increases, respectively, explaining the phase 
agreement with DSMC seen in Fig.\ \ref{fig:timetrace_lambda10}. 
Performance of WSINDy-PI models is consistent across $\lambda$ values (Fig.\ \ref{fig:PIres}).

A primary benefit of our WFEL methodology is the explicit nature of the identified models, from which physical origins can be sought. For instance, $\pmb{F}_{\text{forcing}}$ might be identified with flow velocity corrections arising from vortex diffusion \cite{Alder70_PRA, Cohen93_PA, Han18_JCP}. In particular, inverse cubic terms $\sigma_z^{-3}$, $(\sigma_\perp\sigma_z^2)^{-1}$, $(\sigma_\perp^2\sigma_z)^{-1}$, $\sigma_\perp^{-3}$ that arise (e.g.\ SI Tab. 9 with $\lambda=8$ WSINDy-PI model) can be derived with the addition of a vortical flow field  $U_i(\boldsymbol{r}, t) = \sum_{jk}\varepsilon_{ijk}\alpha_j\frac{r_k}{\sigma_k^2}$ which preserves mass and fluid momentum conservation (SI Sec. 1.1). 
Above, $\varepsilon_{ijk}$ is the Levi-Civita tensor and $\alpha_j$ are model coefficients. We leave further analysis of this mechanism and possibilities for its experimental observation to a future work. Finally, given that WSINDy takes only seconds on a modern laptop for model identification, this work paves the way for ROM discovery and calibration from real experimental data, which will be immediately useful for experimental design and optimization of evaporative cooling.

The authors would like to thank J. L. Bohn for valuable discussions and a thorough reading of this Letter. 
This material is based upon work supported by the National Science Foundation under Grant Number PHY 2317149. 
R.R.W.W. and D.M. contributed equally to this work. 

\bibliography{main_supp.bib}

\onecolumngrid
\clearpage

\begin{center}
{ \Large {\bf Supplementary Information: Physics-guided weak-form discovery of reduced-order models for ultracold anisotropic hydrodynamics} } \vspace{10pt}

{ \large Reuben R. W. Wang$^1$ and Daniel Messenger$^2$ } \vspace{10pt}

{ \normalsize {\it $^1$JILA, NIST, and Department of Physics, University of Colorado, Boulder, Colorado 80309, USA} } 

{ \normalsize {\it $^2$Department of Applied Mathematics, University of Colorado Boulder} }
\end{center}

\setcounter{page}{1}
\setcounter{equation}{0}
\setcounter{table}{0}

\section{ The Ansatz Model \label{app:ansatz_model} }

While detailed in Ref.~\cite{Wang23_PRA2}, we briefly outline the derivation of 
the ansatz model from the fluid equations for completeness of this document. 
The hydrodynamic equations of motion for a nondegenerate gas are \cite{Fetter03_Dover}:
\begin{subequations} \label{eq:continuum_conservation_laws_trapped}
\begin{align} \label{eq:continuity}
    \frac{ \partial \rho }{ \partial t } 
    +
    \sum_j \partial_j \left( { \rho U_j } \right) 
    &= 
    0, \\ \label{eq:Navier_Stokes}
    \frac{\partial}{\partial t} \left( \rho U_i \right) 
    +
    \sum_j \partial_j \left( \rho U_j U_i \right) 
    &= 
    -
    \partial_i 
    \left( n k_B T \right)
    - 
    n \partial_i V(\boldsymbol{r}) 
    +
    \sum_{j, k, \ell} 
    \partial_j 
    \left( \mu_{ i j k \ell } \partial_{\ell} U_k \right), \\ \label{eq:temperature_balance}
    \frac{\partial}{\partial t} (\rho T) 
    +
    \sum_j \partial_j \left( \rho T U_j \right) 
    &= 
    -
    \frac{ 2 }{ 3 }
    \rho T
    \sum_i \partial_i U_i 
    +
    \frac{ 2 m }{ 3 k_B } 
    \sum_{i, j, k, \ell} 
    ( \partial_j U_i )
    \mu_{ i j k \ell }
    ( \partial_{\ell} U_k )
    +
    \frac{ 2 m }{ 3 k_B } 
    \sum_{i, j} 
    \partial_i \left( \kappa_{i j} \partial_j T \right),
\end{align}
\end{subequations}
written in terms of the molecular velocity averaged field variables of mass density, flow velocity and temperature:
\begin{subequations}
\begin{align}
    \rho(\boldsymbol{r}, t) 
    &= 
    m n(\boldsymbol{r}, t) 
    =
    \int d^3 v f(\boldsymbol{r}, \boldsymbol{v}, t) 
    m, \\
    \boldsymbol{U}(\boldsymbol{r}, t) 
    &= 
    \frac{1}{n(\boldsymbol{r}, t)} 
    \int d^3 v f(\boldsymbol{r}, \boldsymbol{v}, t) 
    \boldsymbol{v}, \\
    T(\boldsymbol{r}, t) 
    &= 
    \frac{ 2 }{ 3 n(\boldsymbol{r}, t) k_B } 
    \int d^3 v f(\boldsymbol{r}, \boldsymbol{v}, t) 
    \frac{1}{2} m \boldsymbol{u}^2, \label{eq:local_temperature}
\end{align}
\end{subequations}
respectively. 
Above, $f(\boldsymbol{r}, \boldsymbol{v}, t)$ denotes the molecular phase space distribution and $\boldsymbol{u}(\boldsymbol{r}) = \boldsymbol{v} - \boldsymbol{U}(\boldsymbol{r})$ is the comoving molecular velocity, relative to the frame of fluid flow. 
Of concern to this study are molecular gases confined within a deep optical dipole trap \cite{Grimm00_AAMOP}, well approximated by a cylindrically symmetric harmonic potential $V(\boldsymbol{r})$. Such external confinement leads to soft boundary conditions, conceivably complicating the fluid dynamics via the fluid-gaseous couplings. 
Appropriate treatment of transport on the molecular level is thus paramount to an accurate description of the continuum dynamics, most readily included via the transport tensors of thermal conductivity $\kappa_{i j}$, and viscosity $\mu_{i j k \ell}$.

The interplay of dilute and hydrodynamic regions complicates reduced-order modeling of the dynamics, though Ref.~\cite{Wang23_PRA2} has attempted a first pass by employing a fluctuating Gaussian density to derive effective equations of motion. Although requiring several levels of approximation, the model of Ref.~\cite{Wang23_PRA2} showed favorable agreement to data generated from full Monte Carlo simulations, making it a useful benchmark for the machine-assisted learning methods applied here. 
That effective model treats the effects of thermal transport as negligible, so we omit any further explicit references to the thermal conductivity.
At nondegenerate temperatures, the thermal energy is expected to greatly exceed the average dipolar mean-field energy per particle, leading to transport tensors that predominantly arise from two-body collisions. 
A collision-induced viscosity is approximated to first-order in the Chapman-Enskog expansion by \cite{Chapman90_CUP}
\begin{align} \label{eq:viscosity_tensor} 
    \boldsymbol{\mu} 
    &=
    -\frac{ 2 }{ \beta } \left( \frac{ n }{ m \beta } \right)^2 
    \left(
    \int d^3 u \boldsymbol{W}(\boldsymbol{u}) \otimes C[ f_0 \boldsymbol{W} ]
    \right)^{-1},
\end{align}
where $\beta = ( k_B T )^{-1}$ is the usual inverse temperature and $\boldsymbol{W} = \boldsymbol{u} \boldsymbol{u}^T - \frac{ 1 }{ 3 } \boldsymbol{u}^2 \boldsymbol{I}$ is a rank-2 comoving velocity tensor with the identity matrix $\boldsymbol{I}$. 
The collision integral
\begin{align} \label{eq:collision_integral}
    C[ f_0 \boldsymbol{W} ]
    &=
    \int d^3 u_1 
    | { \boldsymbol{u} - \boldsymbol{u}_1 } |
    f_0(\boldsymbol{u}) f_0(\boldsymbol{u}_1)
    \int d\Omega'
    \frac{ d\sigma }{ d\Omega' }
    \Delta\boldsymbol{W},
\end{align}
with $\Delta\boldsymbol{W} = \boldsymbol{W}' + \boldsymbol{W}'_1 - \boldsymbol{W} - \boldsymbol{W}_1$ and primes denoting post-collision quantities, 
are evaluated with the Maxwell-Boltzmann equilibrium phase space distribution function $f_0(\boldsymbol{u})$ \cite{Reif09_Waveland}. 
The symbol $\otimes$ denotes a dyadic product which takes two tensors of rank $N_1$ and $N_2$, and forms a tensor of rank $N_1 + N_2$ (e.g. $A_{i j} \otimes B_{k \ell} = C_{i j k \ell}$).  Notably, the differential cross section $d\sigma / d\Omega$, resultant from ultracold field-polarized molecular scattering is one necessarily described by quantum mechanics \cite{Bohn09_NJP}, and has an analytic form in the low energy regime \cite{Bohn14_PRA}. 
Refs.~\cite{Wang22_PRA, Wang22_PRA2} have used this cross section to derive analytic expressions for $\boldsymbol{\mu}$ and $\boldsymbol{\kappa}$.

In general, direct solutions to Eq.~(\ref{eq:continuum_conservation_laws_trapped}) require involved numerical solvers to obtain them. We therefore aim to construct a reduced order model by limiting the initial excitations to ones induced by increasing $\omega_z$ instantaneously, restricting the dynamics to be effectively encompassed by a small set of collective observables.
One way of obtaining such a reduced order model is with a suitably chosen ansatz and ensemble averaging, a task we briefly describe below.

\subsection{ Fluctuating Gaussian ansatz: background from [Wang \& Bohn, PRA 108, 013322 (2023)] } At equilibrium, the density distribution enforced by a harmonic trap follows
\begin{align} \label{eq:equilibrium_density_distribution}
    \rho_0(\boldsymbol{r})
    &=
    \frac{ m N }{ Z }
    \exp\left(
    -\frac{ V(\boldsymbol{r}) }{ k_B T_0 } 
    \right),
\end{align}
where $Z = \int d^3 r {\rm e}^{ -\frac{ V(\boldsymbol{r}) }{ k_B T_0 } }$ gives the appropriate normalization and $N$ is the number of molecules. 
The out-of-equilibrium excitations primarily induced by a trap quench of $\omega_z$, are oscillatory modes in the Gaussian density distribution widths, motivating an ansatz of the form
\begin{align} \label{eq:Gaussian_density_ansatz}
    \rho(\boldsymbol{r}, t)
    =
    m N
    \prod_{i=1}^3
    \frac{ 1 }{ \sqrt{ 2 \pi \sigma_i^2(t) } }
    \exp\left(
    -\frac{ r_i^2 }{ 2 \sigma_i^2(t) }
    \right),
\end{align}
where $\sigma_i(t)$ is the distribution widths along each axis $i$ that we allow to vary in time.
These widths will be taken as the observables of interest to the gas dynamics, obtainable in ultracold experiments with in-situ imaging of the sample. Enforcing the continuity equation \eqref{eq:continuity} on \eqref{eq:Gaussian_density_ansatz} yields a solution for the flow field
\begin{equation}\label{eq:ansatz_flow_field}
U_i(\rpf) = \left(\frac{\dot{\sigma_i}(t)}{\sigma_i(t)}\right)r_i.
\end{equation}
Under this approximation, Ref.~\cite{Wang23_PRA2} derives a set of coupled differential equations that govern the time evolution of each $\sigma_i$ from the momentum balance \eqref{eq:Navier_Stokes} and average temperature \eqref{eq:local_temperature}: 
\begin{align} \label{eq:widths_ODE_explicit}
    \ddot{\sigma}_i(t) 
    & +
    \omega_i^2 \sigma_i(t)
    +
    \frac{ 1 }{ 3 \sigma_i(t) }
    \sum_j
    \left[
    \omega_j^2 \sigma_j^2(t) 
    +
    \dot{\sigma}_j^2(t)
    \right]
    -
    \frac{ 2 k_B T_0 }{ m \sigma_i(t) } 
    \approx  
    - \frac{ 2 }{ 5 }
    \frac{ {\cal V}_{\rm hy} }{ N m } 
    \sum_{j} 
    \frac{ \mu_{i i j j}( T(t) ) }{ \sigma_i(t) }
    \frac{ \dot{\sigma}_{j}(t) }{ \sigma_{j}(t) },
\end{align}
where the relevant viscosity matrix elements are written in terms of a unit-free matrix
\begin{align} \label{eq:viscosity_matrix}
    M_{i j}(\Theta)
    &\equiv
    \frac{ \mu_{i i j j}(T; \Theta) }{ \mu_0(T) } \\
    &= 
    \frac{ 1 }{ 512 }
    \begin{pmatrix}
         117 \cos(4\Theta) + 84 \cos(2\Theta) + 415  & 
        -28 ( 3 \cos(2\Theta) + 11 ) & 
        -( 117 \cos(4\Theta) + 107 ) \\[2.5pt]
        -28 ( 3 \cos(2\Theta) + 11 ) & 
        616 & 
        28 ( 3 \cos(2\Theta) - 11 ) \\[2.5pt]
        -( 117 \cos(4\Theta) + 107 ) & 
        28 ( 3 \cos(2\Theta) - 11 ) &
        117 \cos(4\Theta) - 84 \cos(2\Theta) + 415
    \end{pmatrix}, \nonumber
\end{align}
with the isotropic viscosity coefficient given by  
\begin{align} 
    \mu_0(T)
    &=  
    \frac{ 5 }{ 16 a_d^2 } \sqrt{ \frac{ m k_B T }{ \pi } }.
\end{align}
See SI.~\ref{sec:flowfieldcorrections} for more further details on deriving Eq.~(\ref{eq:widths_ODE_explicit}). 
Above, $\Theta = \cos^{-1}\hat{\boldsymbol{{\cal E}}} \cdot \hat{\boldsymbol{z}}$ parameterizes the dipole orientation axis $\hat{\boldsymbol{{\cal E}}}$, which we restrict to lie within the $x,z$-plane.  
The hydrodynamic volume ${\cal V}_{\rm hy}$ is a coarse-graining parameter necessary to formulate the ansatz model, 
identified as the effective volume over which viscosity serves as a good description of dissipative processes.  
In \cite{Wang23_PRA2}, ${\cal V}_{\rm hy}$ was empirically determined to have the approximate functional form 
\begin{align} \label{eq:hydrodynamic_volume_functional_form}
    {\cal V}_{\rm hy}( \lambda, N, \Theta )
    &\approx
    \frac{ 4 \pi }{ 3 }
    \left(
    \frac{ 6 k_B T_0 }{ m \omega_{\perp}^2 }
    \right)^{3/2} 
    \frac{ 1 }{ \sqrt{ \lambda } }
    \left[
    2.21
    +  
    0.67
    \left( 
    1 
    +
    0.26
    \frac{ \sigma_{\rm coll}(\Theta) }{ \overline{\sigma}_{\rm coll} } 
    \right)
    \frac{ N }{ 10^5 }
    \right], 
\end{align}
for cigar geometries ($\lambda < 1$), where $\lambda = ( \omega_z / \omega_{\perp} )^2$ is the aspect ratio of the cylindrically symmetric trap and 
\begin{align} \label{eq:postcollision_averaged_cross_section}
    \sigma_{\rm coll}( \Theta )
    =
    a_d^2 \frac{ \pi }{ 3 }
    \big[
    3 
    +
    18 \cos^2 \Theta
    -
    13 \cos^4 \Theta
    \big],
\end{align}
is the elastic total collision cross section with its angular average being $\overline{\sigma}_{\rm coll} = { 32 \pi a_d^2 / 15 }$ \cite{Bohn14_PRA}. 

We consider the particular case of $\Theta = 0$ (i.e. aligning the dipoles along the trap $z$ axis) in this work, so the relevant viscosity tensor and hydrodynamic volume greatly simplify to 
\begin{subequations}
\begin{align}
    \label{eq:visc_tensor}\mu_{i i j j}(T; 0)
    &=
    \mu_0(T) M_{i j}(0) 
    =
    \frac{ \mu_0(T) }{ 64 }
    \left(
    \begin{array}{ccc}
     77 & -49 & -28 \\
     -49 & 77 & -28 \\
     -28 & -28 & 56 \\
    \end{array}
    \right), \\
    {\cal V}_{\rm hy}( \lambda, N, 0 )
    &\approx
    \frac{ 4 \pi }{ 3 }
    \left(
    \frac{ 6 k_B T_0 }{ m \omega_{\perp}^2 }
    \right)^{3/2} 
    \frac{ 1 }{ \sqrt{ \lambda } } 
    \left[
    2.21
    +  
    0.89
    \frac{ N }{ 10^5 }
    \right], \nonumber
\end{align}
\end{subequations}
enforcing a strict cylindrical symmetry about $z$ in the system. 
The equations of motion are therefore fully encapsulated by the dynamics of just $\sigma_{\perp}(t)$ and $\sigma_z(t)$ as stated in the main text. 

\subsection{Fluctuating Gaussian ansatz: extended derivation with flow field corrections}\label{sec:flowfieldcorrections}

Identification of physical mechanisms behind terms identified in the WSINDy-PI models leads to consideration of extensions to the ansatz model introduced above. In particular, one can consider perturbations $W_i$ of the flow field \eqref{eq:ansatz_flow_field}. Letting the flow field take the form $U_i = V_i+W_i$ with $V_i$ given by \eqref{eq:ansatz_flow_field}, the continuity equation \eqref{eq:continuity} reduces to 
\begin{align}
    \sum_i\left(\partial_i -\frac{r_i}{\sigma_i^2}\right)\cdot W_i = 0,
\end{align}
which in turn implies $\sum_i\sigma_i^{-2}\lan r_iW_i\ran = 0$ using integration by parts and the decay of $\rho(\boldsymbol{r}, t)$ away from the trap center, where $\lan \cdot \ran = (1/N) \int d^3r (\cdot) n(\rpf,t)$ denotes the spatial average. If we restrict each $W_i$ to satisfy $\lan r_i W_i\ran=0$ individually, then we have $\lan V_i W_i\ran=0$, so that the perturbation $W_i$ is uncorrelated with $V_i$. We will assume this in the simplified derivation below, which leads to equations of motion for the widths $\pmb{\sigma}$ that naturally incorporate such flow field corrections. Additionally, we present a specific flow correction $\bsW$ that satisfies these constraints and leads to terms identified in the WSINDy-PI models.

First, we filter the Navier-Stokes equations \eqref{eq:Navier_Stokes} by multiplying each equation by $r_i$ and integrating over space. Using integration by parts together with the rapidly decaying density assumption, the following integral quantities transform according to 
\begin{subequations}
\begin{align}
\int d^3r\, r_i \partial_t(\rho U_i)&= \partial_t\int d^3r\, r_i\rho U_i\\
\int d^3r\, r_i\sum_j \partial_j \left( \rho U_j U_i \right)
    &= \int d^3r\, \sum_j \left[\partial_j \left( r_i\rho U_j U_i \right) - \delta_{ij}\rho U_j U_i\right] =-\int d^3r\,\rho (U_i)^2\\
    \int d^3r\, r_i\partial_i 
    \left( n k_B T \right) &= -\int d^3r\, nk_BT\\
    \int d^3r\, r_i n \partial_i V(\boldsymbol{r}) &= \omega_i^2\int d^3r\, r_i^2 \rho = mN \omega_i^2 \sigma_i^2\\
     \int d^3r\, r_i\sum_{j, k, \ell} 
    \partial_j 
    \left( \mu_{ i j k \ell } \partial_{\ell} U_k \right) &= -\int d^3r\, \sum_{j, k, \ell} 
    \delta_{ij} 
    \left( \mu_{ i j k \ell } \partial_{\ell} U_k \right) = -\int d^3r\, \sum_{k, \ell} 
    \left( \mu_{ i i k \ell } \partial_{\ell} U_k \right)
\end{align}
\end{subequations}
With this in mind, we need only calculate the quantities on the right-hand sides in terms of $\bsV$ and $\bsW$, assumed to be uncorrelated:
\begin{subequations}
\begin{align}
    \partial_t\int d^3r\,r_i\rho U_i
    &=
    \partial_t\int d^3r\,r_i\rho V_i
    =
    mN\sigma_i\ddot{\sigma}_i + mN(\dot{\sigma}_i)^2, \\
    -\int d^3r\,\rho (U_i)^2 
    &=
    -\int d^3r\,\rho (V_i)^2- \int d^3r\,\rho (W_i)^2 
    =
    -mN(\dot{\sigma}_i)^2 -mN\lan W_i^2\ran, \\
    -\int d^3r\, nk_BT 
    &= -N k_B\lan T\ran, \\
    -\int d^3r\, \sum_{k, \ell} 
    \left( \mu_{ i i k \ell } \partial_{\ell} U_k \right)  
    &=
    -N \lan 
    \sum_{k, \ell} 
    \left( \frac{ \mu_{ i i k \ell } }{ n }
    \partial_{\ell} U_k \right)
    \ran 
    =
    -N \sum_{k} 
    \lan \frac{ \mu_{ i i kk } }{ n } \ran 
    \frac{\dot{\sigma}_k}{\sigma_k}
    -
    N \lan
    \sum_{k, \ell} 
    \left( 
    \frac{\mu_{ i i k \ell }}{n}
    \partial_{\ell} W_k 
    \right)
    \ran.
\end{align}
\end{subequations}
Putting this all together, the NS equations become, after dividing through by $mN\sigma_i$,
\begin{align}
    \ddot{\sigma}_i - \frac{1}{\sigma_i}\lan W_i^2\ran +\omega_i^2 \sigma_i - \frac{ k_B}{m\sigma_i}\lan T\ran 
    &=
    -\frac{1}{m} 
    \sum_{k} 
    \lan 
    \frac{\mu_{ i i kk }}{n} 
    \ran 
    \frac{\dot{\sigma}_k}{\sigma_i\sigma_k}
    -
    \frac{1}{m\sigma_i}
    \lan 
    \sum_{k, \ell} 
    \left(
    \frac{ \mu_{ i i k \ell } }{ n } 
    \partial_{\ell} W_k 
    \right)\ran.
\end{align}
For the average temperature, we have by definition that 
\begin{align}
    \frac{3}{2}nk_BT 
    &=
    \frac{m}{2}\sum_{i=1}^3 
    \left(
    \int d^3v f(r,v,t)v_i^2 - \frac{1}{2}\rho U_i^2
    \right),
\end{align}
and 
\begin{align}
    E_{\rm total} 
    &=
    \frac{m}{2}\sum_{i=1}^3 \left(\omega_i^2\lan r_i^2\ran+\frac{1}{N}\int d^3rd^3v f(r,v,t)v_i^2\right) =3k_BT_0,
\end{align}
with $T_0$ the equilibrium temperature (assumed constant in space), so that 
\begin{align}
    3k_BT_0 
    &=
    \frac{m}{2}\sum_{i=1}^3\omega_i^2\sigma_i^2 +\frac{1}{2N}\sum_{i=1}^3\int d^3r\rho(U_i)^2+ \frac{3}{2}k_B\lan T\ran,
\end{align}
or 
\begin{align}
    \frac{k_B}{m}\lan T\ran 
    &=
    \frac{2k_BT_0}{m} - \frac{1}{3}\sum_{i=1}^3 \omega_i^2\sigma_i^2 - \frac{1}{3}\sum_{i=1}^3\left((\dot{\sigma}_i)^2 + \lan W_i^2\ran\right).
\end{align}
Substituting this in for $\lan T\ran$ above, we get 
\begin{subequations}
\begin{align} \label{eq:ans_rederiv} 
    \ddot{\sigma}_i(t) 
    +
    \omega_i^2 \sigma_i(t)
    +
    \frac{ 1 }{ 3 \sigma_i(t) }
    \sum_j
    &\left[
    \omega_j^2 \sigma_j^2(t) 
    +
    \dot{\sigma}_j^2(t)
    \right]
    -
    \frac{ 2 k_B T_0 }{ m \sigma_i(t) }  = -\frac{1}{m} \sum_{k} \lan \frac{ \mu_{ i i kk } }{ n } \ran \frac{\dot{\sigma}_k}{\sigma_i\sigma_k} \\ 
    \label{eq:ans_rederiv_corr}
    &- 
    \frac{1}{\sigma_i} 
    \lan
    W_i^2 
    \ran
    -
    \frac{1}{3}\sum_{j=1}^3 \frac{1}{\sigma_i}
    \lan W_j^2\ran  
    -
    \frac{1}{m\sigma_i } \sum_{k, \ell, p} 
    \lan \frac{ \mu_{ i i k \ell } }{ n } \partial_\ell W_k
    \ran
\end{align}
\end{subequations}
where we see that \eqref{eq:ans_rederiv} contains the original ansatz (resulting from only $V_i$) while \eqref{eq:ans_rederiv_corr} contains the effects of velocity corrections $W_i$. Using the particular form 
\begin{equation}\label{eq:vorticalflow}
W_i = \sum_{j, k} \varepsilon_{i j k}
    \frac{ \alpha_j}{ \sigma_k^2}r_k
\end{equation}
for $\alpha_j$ undetermined (and possibly time-dependent), which is consistent with the continuity equation and additional satisfies $\lan V_i W_i\ran=0$, we arrive at
\begin{subequations}
\begin{align}
 \ddot{\sigma}_i(t) 
    +
    \omega_i^2 \sigma_i(t)
    +
    \frac{ 1 }{ 3 \sigma_i(t) }
    \sum_j
    &\left[
    \omega_j^2 \sigma_j^2(t) 
    +
    \dot{\sigma}_j^2(t)
    \right]
    -
    \frac{ 2 k_B T_0 }{ m \sigma_i(t) }  
    =
    -\frac{1}{m} \sum_{k} \lan \frac{ \mu_{ i i kk } }{ \nu } \ran \frac{\dot{\sigma}_k}{\sigma_i\sigma_k} \\
    &-\sum_{j,k} |\varepsilon_{ijk}|
    \frac{\alpha_j^2}{\sigma_i\sigma_k^2}
    -
    \frac{1}{3}\sum_{j,l,m}
    |\varepsilon_{j\ell m}|
    \frac{\alpha_\ell^2}{\sigma_i\sigma_m^2} 
    -
    \frac{1}{m} \sum_{k, \ell, p} 
    \lan 
    \frac{ \mu_{ i i k \ell } }{ n } 
    \ran \varepsilon_{kp\ell} 
    \frac{\alpha_p}{\sigma_i \sigma_\ell^2}
\end{align}
\end{subequations}
which provides one possible explanation for the inverse cubic terms appearing in the WSINDy-PI models.

\section{WSINDy algorithmic details}

The quasi-matrix libraries in Tab.~1 of the main text are given a matrix representation when evaluated on the training data.  That is, given data $\Xbf \in \Rbb^{M\times d}$,  $\Lbb(\Xbf)$ takes the matrix form
\begin{align}
    \Lbb(\Xbf)  
    =
    \begin{bmatrix} 
    \vert & & \vert \\
    f_1(\Xbf) & \cdots &f_J(\Xbf) \\
    \vert & & \vert
    \end{bmatrix} \in \Rbb^{M\times J}.
\end{align}
For library terms that are nonlinear in the first time derivative, we approximate $\dot{\sigma}_\perp$ and $\dot{\sigma}_z$ using 2nd-order centered finite differences. All other appearances of time derivatives are computed in weak form by placing the time derivative onto the test function via integrating by parts.

We enforce the weak form on a basis of test functions 
\[\phi_k(t) = \varphi(t-t_k;m_t\Delta t,\eta)\]
where $\varphi$ is the $C^\infty_c$ bump function
\[\varphi(t;a,\eta) = \exp\left(-\frac{\eta}{1-(t/a)^2}\right)\ind{(-a,a)}(t)\]
and $\Delta t$ is the sampling rate of the data, which is uniformly distributed in time. The parameters $\eta,m_t$ are solved for in each $\sigma_i$ equation using the method outlined in \cite[Appendix A]{messenger2021weak}, by finding a cornerpoint $k^*_i$ in the cumulative sum of the power spectrum of $\sigma_i$, and assigning $k^*_i$ to be two standard deviations into the tail of the power spectrum of $\varphi(t;a,\eta)$, the latter viewed as a probability distribution over frequency space. From $m_t$, the query points $\{t_k\}_{k=1}^K$ are chosen equally spaced such that $t_k-t_{k-1}=\lfloor m_t/4\rfloor$. 

\subsection{Rescaling}\label{app:rescaling}

In dimensional variables with SI-units, fluctuations in the widths $(\sigma_i)_{i=1}^3$ are $\CalO(10^{-5})$, while coefficients in the ansatz model 
can range from $\CalO(10^{-10})$ to $\CalO(10^{6})$, necessitating rescaling of the data, in the form 
\begin{align}
    \tilde{\sigma}_i 
    &=
    \sigma_i / \gamma_\sigma, \quad
    \tilde{t} 
    =
    t/\gamma_t, \quad
    \tilde{m}
    =
    m / \gamma_m, 
\end{align}
for some characteristic scales $\gamma_\sigma, \gamma_t$ and $\gamma_m$ in order to identify dominant terms using sparse regression. In the na\"ive library approach, $\gamma_\sigma$, the lengthscale, is chosen to enforce that the aggregate widths data $(\sigma_i)_{i=1}^3$ evaluated at the time grid $\tbf$ has unit absolute mean. The timescale $\gamma_t$ is chosen to be the period of the dominant slow mode of the dynamics, computed as the average difference between local maxima in the $\sigma_z$-coordinate using \texttt{findpeaks} in MATLAB.
The mass scale is simply the molecular mass $\gamma_m = m$. 

For the ansatz verification (AV) and physics-informed (PI) library approaches, the length scale $\gamma_\sigma$ is chosen such that the equipartitioned thermal energy matches the potential energy of the harmonic trap. Meanwhile, $\gamma_t$ is chosen to enforce that the geometric mean of the natural trap oscillation frequencies is unity. Similarly, $\gamma_m = m$ once more. Under this rescaling, terms significant to the dynamics yield coefficients that remain similar in magnitude, while superfluous terms have much smaller coefficients that are easily thresholded.

\subsection{Sparse regression}\label{app:MSTLS}

To arrive at a sparse weight matrix $\what\approx \wstar$, an approximate solution is sought for the $\ell_0$-regularized least-squares problem
\begin{equation}\label{eq:stls}
\min_{\wbf\in\Rbb^{J\times d}}\left\Vert \Gbf \wbf-\bbf \right\Vert_F^2 +\nu^2\left\Vert\wbf\right\Vert_0 
\end{equation}
where $\|\cdot\|_F$ denotes the Frobenius norm and $\|\wbf\|_0$ is the number of nonzero elements of $\wbf$, referred to as the $\ell_0$-pseudonorm. The WSINDy algorithm involves solving \eqref{eq:stls} using the Modified Sequential Thresholding Least-Squares algorithm (MSTLS) \cite{messenger2021weak}, which employs alternating hard thresholding steps and least-squares solves, together with a grid search for the sparsity threshold $\nu$. In all cases we use the MSTLS algorithm with a grid of 300 equally log-spaced values $\pmb{\nu}$ for the sparsity threshold $\nu$, ranging from $10^{-8}$ to $10^{-1}$.

\section{ Supplementary figures and tables \label{app:supp_figures} }

\begin{table}[ht]
\resizebox{\textwidth}{!}{
\begin{tabular}{l|cccccccc}
& $\lambda=0.13$ & $\lambda=0.2$ & $\lambda=0.32$ & $\lambda=0.5$ & $\lambda=2$ & $\lambda=3.2$ & $\lambda=5$ & $\lambda=8$ \\ 
\hline 
$\sigma_\perp^{-1}$ & 2.156e-04 & 2.156e-04 & 2.156e-04 & 2.156e-04 & 2.156e-04 & 2.156e-04 & 2.156e-04 & 2.156e-04 \\ 
\textcolor{white}{1} & 2.158e-04 & 2.158e-04 & 2.157e-04 & 2.158e-04 & 2.157e-04 & 2.158e-04 & 2.158e-04 & 2.157e-04 \\ 
$\dot{\sigma}_\perp\sigma_\perp^{-2}$ & -1.689e-08 & -1.689e-08 & -1.689e-08 & -1.689e-08 & -1.689e-08 & -1.689e-08 & -1.689e-08 & -1.689e-08 \\ 
\textcolor{white}{2} & -3.621e-08 & -3.787e-08 & -1.928e-08 & -2.356e-08 & -1.968e-08 & -2.239e-08 & -2.469e-08 & -2.576e-08 \\ 
$\dot{\sigma}_z(\sigma_\perp\sigma_z)^{-1}$ & 1.689e-08 & 1.689e-08 & 1.689e-08 & 1.689e-08 & 1.689e-08 & 1.689e-08 & 1.689e-08 & 1.689e-08 \\ 
\textcolor{white}{3} & 1.113e-08 & 9.819e-09 & 1.483e-08 & 1.392e-08 & 1.285e-08 & 1.144e-08 & 1.193e-08 & 1.039e-08 \\ \hline
$\sigma_z^{-1}$ & 2.156e-04 & 2.156e-04 & 2.156e-04 & 2.156e-04 & 2.156e-04 & 2.156e-04 & 2.156e-04 & 2.156e-04 \\ 
\textcolor{white}{4} & 2.161e-04 & 2.159e-04 & 2.161e-04 & 2.160e-04 & 2.161e-04 & 2.160e-04 & 2.161e-04 & 2.161e-04 \\ 
$\dot{\sigma}_\perp(\sigma_\perp\sigma_z)^{-1}$ & 3.378e-08 & 3.378e-08 & 3.378e-08 & 3.378e-08 & 3.378e-08 & 3.378e-08 & 3.378e-08 & 3.378e-08 \\ 
\textcolor{white}{5} & 7.585e-08 & 5.206e-08 & 3.617e-08 & 3.943e-08 & 2.888e-08 & 3.155e-08 & 3.109e-08 & 3.558e-08 \\ 
$\dot{\sigma}_z\sigma_z^{-2}$ & -3.378e-08 & -3.378e-08 & -3.378e-08 & -3.378e-08 & -3.378e-08 & -3.378e-08 & -3.378e-08 & -3.378e-08 \\ 
\textcolor{white}{6} & -4.762e-08 & -4.482e-08 & -4.460e-08 & -4.182e-08 & -3.144e-08 & -2.786e-08 & -2.692e-08 & -2.311e-08 
\end{tabular}
}
\caption{Comparison between Ansatz coefficients (top of each row) and Weak-form approximated coefficients (bottom of each row) while material derivative terms are enforced. The top 3 rows correspond to the equation for $\sigma_\perp$, while the bottom 3 rows correspond to  $\sigma_z$. }
\label{tab:true_coeffs}
\end{table}

In Table \ref{tab:true_coeffs} we include a comparison of coefficient values in the ansatz model (top of each row) and WSINDy-AV (bottom of each row) models over all trap aspect ratios $\lambda$. Coefficients of $\sigma_\perp^{-1}$ and $\sigma_z^{-1}$ agree to four significant digits, indicating that the leading-order equation of state (EOS) is well specified in the ansatz model. All other coefficients show a deviation between the two models, indicating that a constant $\CalV_\text{hy}$ assumption may be too rigid (see Figure 1 of the main text and supporting descriptions).

In Tables \ref{tab:lambda1}-\ref{tab:lambda10} we include the WSINDy-PI learned equations along with visualizations of trajectories simulated from these learned models in comparison to the ansatz model and the DSMC data, with Table \ref{tab:lambda10} a continuation of Figures 3-4 in the main text. The visualizations include simulations of the WSINDy-PI and ansatz models up to twice the length of the DSMC data used in training in order to depict differences in the long-term behavior of the two models. In particular, for cigar-shaped traps (Tables \ref{tab:lambda1}-\ref{tab:lambda4}), WSINDy-PI models appear to more accurately capture the long-term damped oscillations. Above each figure the relative error (defined in the main text) between the DSMC data and ansatz (Ans) or WSINDy (WS) models is listed, showing an approximate reduction by an order of magnitude from the ansatz to WSINDy model. Each table includes terms that were selected by the WSINDy algorithm, their respective coefficient values (such that if $f_j$ is an included term in the $i$th equation with coefficient $\what_{ji}$ listed in the table, then $\ddot{\sigma}_i = \cdots + \what_{ji}f_j +\dots$), and their respective term magnitudes. The term magnitude of the $j$th term in the $i$th equation is given by $|\what_{ji}| \times |\lan \Gbf_j,\bbf_i\ran|/\|\bbf_i\|_2^2$. These term magnitudes provide a measure of relative importance of the terms to the weak-form dynamics, with a value near 1 indicating that the term is dominant. For example, the term magnitudes for $\sigma_\perp$ and $\sigma_z$ are $\CalO(1)$ in each $\ddot{\sigma}_\perp$ and $\ddot{\sigma}_z$ equation, respectively, reflecting that the overall harmonic oscillations are dominant. 

\begin{table}[ht]
\begin{tabular}{@{}c@{}c@{}}
\includegraphics[trim={15 3 35 5},clip,width=0.5\textwidth]{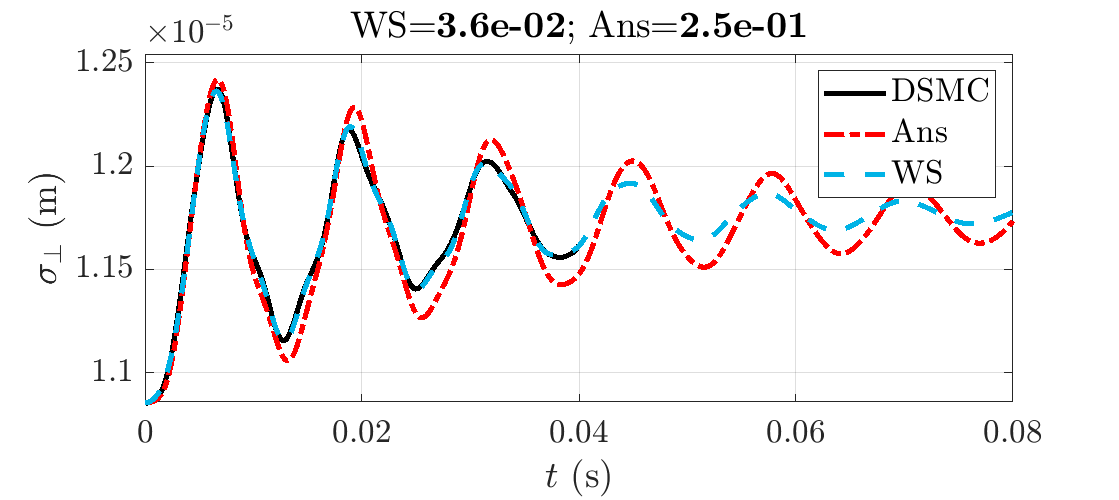} &
\includegraphics[trim={15 3 35 5},clip,width=0.5\textwidth]{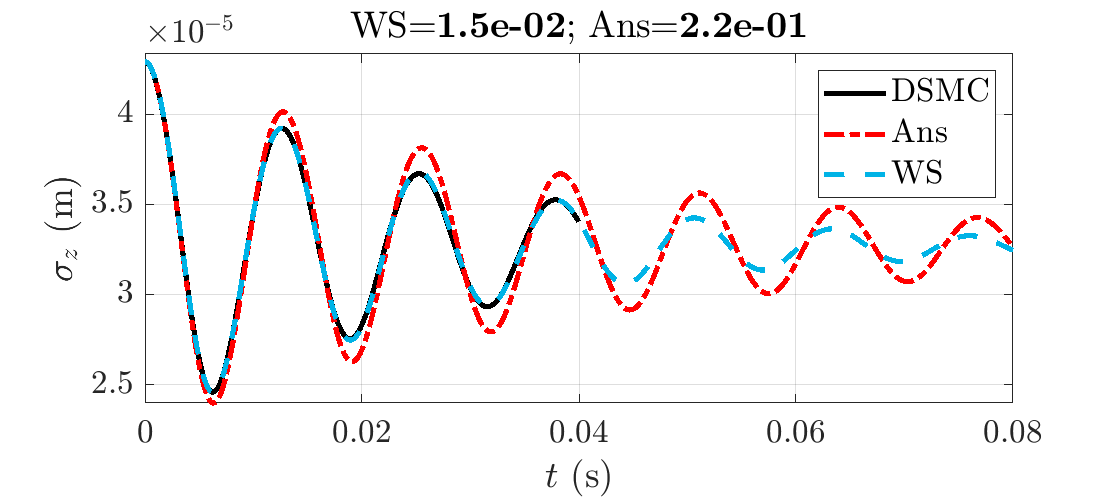}
\end{tabular}

\begin{tabular}{@{}c@{}|c@{}}
\begin{minipage}{0.5\textwidth}
\resizebox{\textwidth}{!}{
\begin{tabular}{p{0.3cm}|p{2cm}|p{0.4cm}|p{1.6cm}|p{1.6cm}|p{1.8cm}}
 & term & AE & Ansatz Coeff & Learned Coeff & term mag \\ \hline
A & $\sigma_\perp$ & 1 & -1.30e+06 & -1.30e+06 & 1.76e+00 \\
& $\sigma_\perp^{-1}$ & 1 & 2.16e-04 & 2.16e-04 & 4.29e+00 \\
& $\sigma_\perp^{-1}\sigma_z^2$ & 1 & -3.38e+04 & -3.38e+04 & 4.85e+00 \\
& $\sigma_\perp^{-1}\dot{\sigma_\perp}^2$ & 1 & -6.67e-01 & -6.67e-01 & 2.24e-04 \\
& $\dot{\sigma}_\perp\sigma_\perp^{-2}$ & 0 & -1.69e-08 & 8.97e-08 & 1.31e-02 \\
& $\sigma_\perp^{-1}\dot{\sigma_z}^2$ & 1 & -3.33e-01 & -3.33e-01 & 3.19e-01 \\
& $\dot{\sigma}_z\sigma_\perp^{-1}\sigma_z^{-1}$ & 0 & 1.69e-08 & 1.20e-08 & 6.18e-02 \\
\hline F & $\sigma_z^{-3}$ & 0 & 0.00e+00 & -2.17e-15 & 1.57e-01 \\
& $\sigma_z^2$ & 0 & 0.00e+00 & 7.75e+07 & 1.16e-01 \\
& $\sigma_z^3$ & 0 & 0.00e+00 & -3.83e+12 & 2.74e-01 \\
& $\sigma_\perp\sigma_z^2$ & 0 & 0.00e+00 & 1.11e+13 & 1.71e-01 \\
\hline V & $\sigma_\perp^{-2}\dot{\sigma}_\perp\dot{\sigma_\perp}^2$ & 0 & 0.00e+00 & 1.17e+00 & 1.08e-02 \\
& $\dot{\sigma}_\perp\sigma_\perp^{-1}\sigma_z^{-1}$ & 0 & 0.00e+00 & 1.11e-06 & 6.70e-02 \\
& $\sigma_\perp^{-1}\sigma_z^{-1}\dot{\sigma}_\perp\dot{\sigma_\perp}^2$ & 0 & 0.00e+00 & -2.98e+00 & 1.32e-02 \\
& $\dot{\sigma}_\perp$ & 0 & -0.00e+00 & -5.17e+03 & 9.33e-03 \\
& $\sigma_\perp^{-1}\sigma_z^{-1}\dot{\sigma}_\perp\dot{\sigma_z}^2$ & 0 & 0.00e+00 & -1.96e-02 & 6.06e-02 \\
& $\dot{\sigma}_z\sigma_z^{-2}$ & 0 & 0.00e+00 & 9.61e-09 & 1.54e-02 \\
& $\sigma_z^{-2}\dot{\sigma}_\perp\dot{\sigma_z}^2$ & 0 & 0.00e+00 & 4.28e-02 & 3.32e-02 \\
& $\sigma_\perp^{-1}\sigma_z\dot{\sigma}_\perp$ & 0 & 0.00e+00 & 5.32e+02 & 6.43e-02 \\
\end{tabular}
}
\end{minipage}&
\begin{minipage}{0.5\textwidth}
\resizebox{\textwidth}{!}{
\begin{tabular}{p{0.3cm}|p{2cm}|p{0.4cm}|p{1.6cm}|p{1.6cm}|p{1.8cm}}
  & term & AE & Ansatz Coeff & Learned Coeff & term mag \\ \hline
A & $\sigma_z$ & 1 & -1.35e+05 & -1.35e+05 & 5.27e-01 \\
& $\sigma_z^{-1}$ & 1 & 2.16e-04 & 2.16e-04 & 8.43e-01 \\
& $\sigma_\perp^2\sigma_z^{-1}$ & 1 & -5.20e+05 & -5.20e+05 & 3.96e-01 \\
& $\sigma_z^{-1}\dot{\sigma_\perp}^2$ & 1 & -6.67e-01 & -6.67e-01 & 3.71e-04 \\
& $\dot{\sigma}_\perp\sigma_\perp^{-1}\sigma_z^{-1}$ & 0 & 3.38e-08 & -8.53e-09 & 7.91e-04 \\
& $\sigma_z^{-1}\dot{\sigma_z}^2$ & 1 & -3.33e-01 & -3.33e-01 & 4.51e-03 \\
& $\dot{\sigma}_z\sigma_z^{-2}$ & 0 & -3.38e-08 & -3.22e-08 & 3.30e-03 \\
\hline F & $\sigma_\perp^{-1}\sigma_z^2$ & 0 & 0.00e+00 & -3.69e+03 & 9.02e-02 \\
& $\sigma_z^{-3}$ & 0 & 0.00e+00 & 2.60e-15 & 3.10e-02 \\
& $\sigma_z^3$ & 0 & 0.00e+00 & 7.63e+12 & 9.70e-02 \\
\hline V & $\dot{\sigma}_\perp\sigma_\perp^{-2}$ & 0 & 0.00e+00 & 3.14e-09 & 6.43e-04 \\
& $\sigma_\perp^{-2}\dot{\sigma}_z\dot{\sigma_\perp}^2$ & 0 & 0.00e+00 & -9.10e-02 & 1.23e-02 \\
& $\sigma_\perp^{-1}\sigma_z^{-1}\dot{\sigma}_z\dot{\sigma_\perp}^2$ & 0 & 0.00e+00 & 2.27e-01 & 1.39e-02 \\
& $\sigma_\perp\sigma_z^{-1}\dot{\sigma}_z$ & 0 & 0.00e+00 & -7.44e+01 & 1.82e-03 \\
& $\dot{\sigma}_z\sigma_\perp^{-1}\sigma_z^{-1}$ & 0 & 0.00e+00 & -1.73e-08 & 2.07e-03 \\
& $\sigma_\perp^{-1}\sigma_z^{-1}\dot{\sigma}_z\dot{\sigma_z}^2$ & 0 & -0.00e+00 & 1.06e-03 & 1.30e-03 \\
& $\sigma_z^{-2}\dot{\sigma}_z\dot{\sigma_z}^2$ & 0 & 0.00e+00 & -2.15e-03 & 2.34e-03 \\
& $\dot{\sigma}_z$ & 0 & 0.00e+00 & 3.49e+01 & 5.09e-05 \\
\end{tabular}}
\end{minipage}
\end{tabular}
\caption{Learned equations for $\ddot{\sigma}_\perp$ (left) and $\ddot{\sigma}_z$ (right) at $\lambda = 0.13$ using WSINDy with the Physics-Informed library approach, together with visualizations of the learned trajectories with respect to DSMC and ansatz results (top). Relative errors (defined in the main text) between the DSMC data and ansatz (Ans) or WSINDy (WS) models is listed above each figure, showing an approximate reduction in error by an order of magnitude from the ansatz to WSINDy model. In the left columns, `A', `F', and `V' indicate terms that from the ansatz, the additional forcing library $\Lbb^{f}$, and the viscous library $\Lbb^{\mu}_i$. The 'AE' column indicates whether the respective term's coefficient was ``ansatz enforced", that is, enforced to take the value derived in the ansatz. The ``term mag" is the magnitude of the relative projection of the respective term onto the dynamics $\ddot{\sigma}_i$ in weak form. For the $j$th term of the $i$th equation, the term magnitude is given by $|\what_{ji}||\lan \Gbf_j,\bbf_i\ran|/\|\bbf_i\|_2^2$, where $(\Gbf,\bbf)$ is the weak form linear system in \eqref{eq:stls}.}
\label{tab:lambda1}
\end{table}

\begin{table}[ht]
\begin{tabular}{@{}c@{}c@{}}
\includegraphics[trim={15 3 35 5},clip,width=0.5\textwidth]{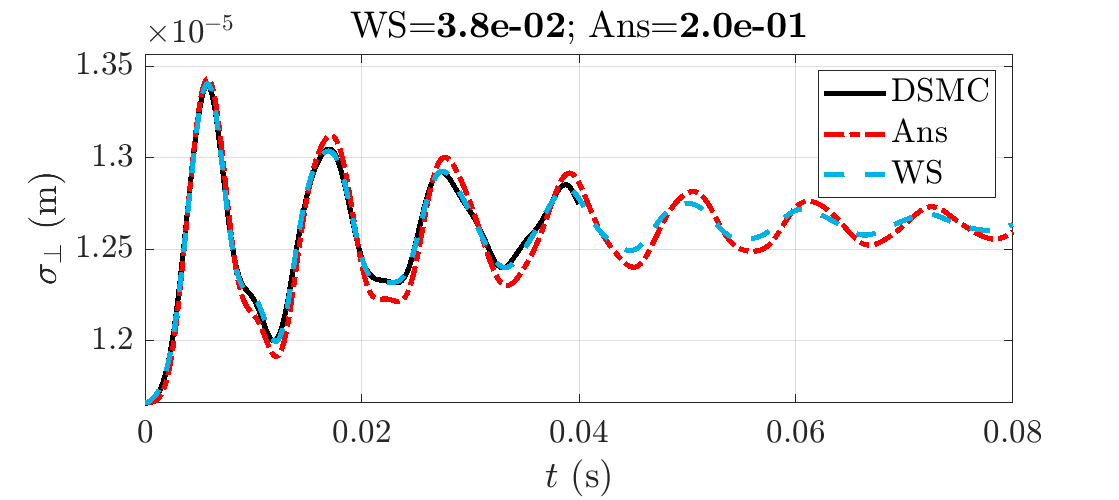} &
\includegraphics[trim={15 3 35 5},clip,width=0.5\textwidth]{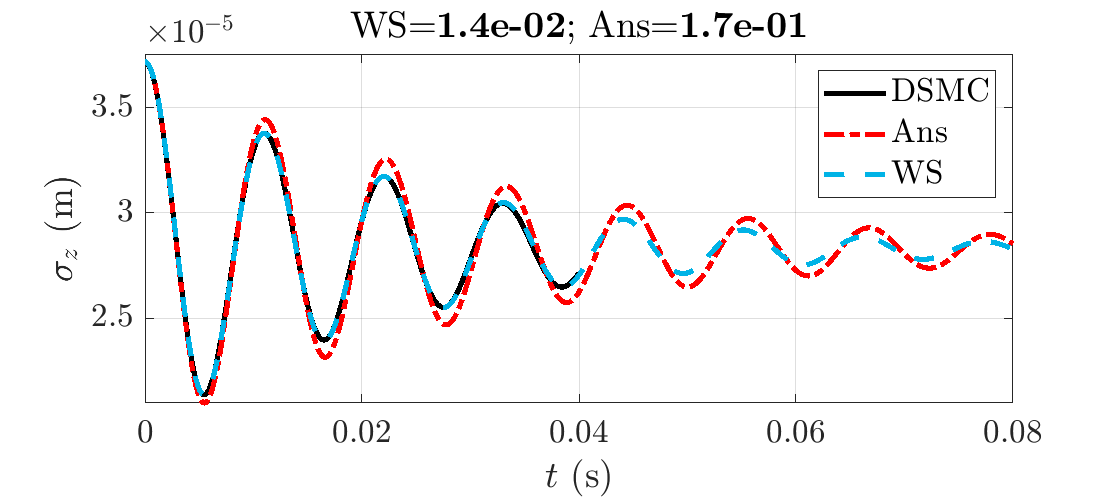}
\end{tabular}

\begin{tabular}{@{}c@{}|c@{}}
\begin{minipage}{0.5\textwidth}
\resizebox{\textwidth}{!}{
\begin{tabular}{p{0.3cm}|p{2cm}|p{0.4cm}|p{1.6cm}|p{1.6cm}|p{1.8cm}}
 & term & AE & Ansatz Coeff & Learned Coeff & term mag \\ \hline
A & $\sigma_\perp$ & 1 & -1.13e+06 & -1.13e+06 & 3.75e+00 \\
& $\sigma_\perp^{-1}$ & 1 & 2.16e-04 & 2.16e-04 & 1.65e-01 \\
& $\sigma_\perp^{-1}\sigma_z^2$ & 1 & -4.50e+04 & -4.50e+04 & 2.66e+00 \\
& $\sigma_\perp^{-1}\dot{\sigma_\perp}^2$ & 1 & -6.67e-01 & -6.67e-01 & 3.16e-04 \\
& $\dot{\sigma}_\perp\sigma_\perp^{-2}$ & 0 & -1.69e-08 & 1.62e-07 & 2.48e-02 \\
& $\sigma_\perp^{-1}\dot{\sigma_z}^2$ & 1 & -3.33e-01 & -3.33e-01 & 3.00e-01 \\
& $\dot{\sigma}_z\sigma_\perp^{-1}\sigma_z^{-1}$ & 0 & 1.69e-08 & 2.75e-08 & 1.31e-01 \\
\hline F & $\sigma_z^{-3}$ & 0 & 0.00e+00 & -4.78e-16 & 4.23e-02 \\
& $\sigma_z^3$ & 0 & 0.00e+00 & 1.75e+12 & 4.87e-02 \\
\hline V & $\sigma_\perp^{-2}\dot{\sigma}_\perp\dot{\sigma_\perp}^2$ & 0 & 0.00e+00 & -3.92e-01 & 5.76e-03 \\
& $\dot{\sigma}_\perp\sigma_\perp^{-1}\sigma_z^{-1}$ & 0 & 0.00e+00 & 1.63e-08 & 7.19e-04 \\
& $\sigma_\perp^{-1}\sigma_z^{-1}\dot{\sigma}_\perp\dot{\sigma_\perp}^2$ & 0 & 0.00e+00 & 5.80e-01 & 4.69e-03 \\
& $\dot{\sigma}_\perp$ & 0 & 0.00e+00 & -9.04e+02 & 5.78e-03 \\
& $\sigma_\perp^{-1}\sigma_z^{-1}\dot{\sigma}_\perp\dot{\sigma_z}^2$ & 0 & 0.00e+00 & 1.48e-03 & 6.65e-03 \\
& $\dot{\sigma}_z\sigma_z^{-2}$ & 0 & -0.00e+00 & -2.77e-08 & 5.31e-02 \\
& $\sigma_z^{-2}\dot{\sigma}_\perp\dot{\sigma_z}^2$ & 0 & 0.00e+00 & 1.30e-03 & 2.23e-03 \\
& $\sigma_\perp^{-1}\sigma_z\dot{\sigma}_\perp$ & 0 & 0.00e+00 & -1.12e+02 & 1.11e-02 \\
\end{tabular}}
\end{minipage}&
\begin{minipage}{0.5\textwidth}
\resizebox{\textwidth}{!}{
\begin{tabular}{p{0.3cm}|p{2cm}|p{0.4cm}|p{1.6cm}|p{1.6cm}|p{1.8cm}}
  & term & AE & Ansatz Coeff & Learned Coeff & term mag \\ \hline
A & $\sigma_z$ & 1 & -1.80e+05 & -1.80e+05 & 3.69e-01 \\
& $\sigma_z^{-1}$ & 1 & 2.16e-04 & 2.16e-04 & 1.10e+00 \\
& $\sigma_\perp^2\sigma_z^{-1}$ & 1 & -4.50e+05 & -4.50e+05 & 4.92e-01 \\
& $\sigma_z^{-1}\dot{\sigma_\perp}^2$ & 1 & -6.67e-01 & -6.67e-01 & 8.31e-04 \\
& $\dot{\sigma}_\perp\sigma_\perp^{-1}\sigma_z^{-1}$ & 0 & 3.38e-08 & -1.19e-07 & 1.60e-02 \\
& $\sigma_z^{-1}\dot{\sigma_z}^2$ & 1 & -3.33e-01 & -3.33e-01 & 4.42e-03 \\
& $\dot{\sigma}_z\sigma_z^{-2}$ & 0 & -3.38e-08 & -2.71e-08 & 3.51e-03 \\
\hline F & $\sigma_\perp^{-1}\sigma_z^2$ & 0 & 0.00e+00 & -5.55e+03 & 7.23e-02 \\
& $\sigma_z^3$ & 0 & 0.00e+00 & 1.20e+13 & 7.84e-02 \\
& $\sigma_\perp\sigma_z^{-2}$ & 0 & 0.00e+00 & 5.70e-06 & 2.61e-02 \\
\hline V & $\dot{\sigma}_\perp\sigma_\perp^{-2}$ & 0 & 0.00e+00 & 5.89e-08 & 1.47e-02 \\
& $\sigma_\perp^{-2}\dot{\sigma}_z\dot{\sigma_\perp}^2$ & 0 & 0.00e+00 & 7.92e-03 & 1.90e-03 \\
& $\sigma_\perp^{-1}\sigma_z^{-1}\dot{\sigma}_z\dot{\sigma_\perp}^2$ & 0 & 0.00e+00 & -3.25e-02 & 4.30e-03 \\
& $\sigma_\perp\sigma_z^{-1}\dot{\sigma}_z$ & 0 & 0.00e+00 & -7.34e+02 & 2.16e-02 \\
& $\dot{\sigma}_z\sigma_\perp^{-1}\sigma_z^{-1}$ & 0 & 0.00e+00 & 1.56e-07 & 2.00e-02 \\
& $\sigma_\perp^{-1}\sigma_z^{-1}\dot{\sigma}_z\dot{\sigma_z}^2$ & 0 & -0.00e+00 & -2.16e-04 & 2.59e-04 \\
& $\sigma_z^{-2}\dot{\sigma}_z\dot{\sigma_z}^2$ & 0 & 0.00e+00 & -4.58e-04 & 5.96e-04 \\
& $\dot{\sigma}_z$ & 0 & 0.00e+00 & -1.43e+02 & 7.37e-04 \\
\end{tabular}}
\end{minipage}
\end{tabular}
\caption{Similar results to Table \ref{tab:lambda1} for $\lambda = 0.2$.}
\label{tab:lambda2}
\end{table}

\begin{table}[ht]
\begin{tabular}{@{}c@{}c@{}}
\includegraphics[trim={15 3 35 5},clip,width=0.5\textwidth]{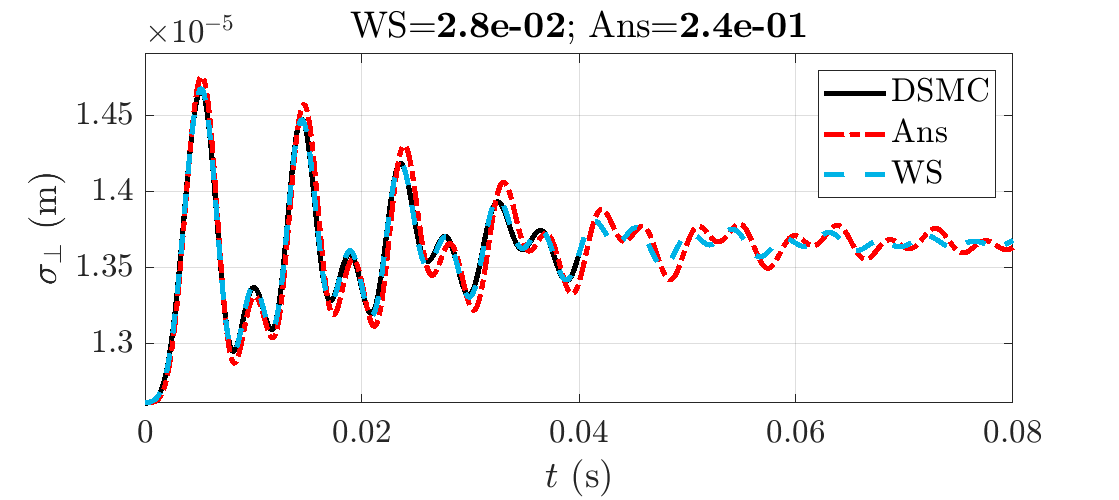} &
\includegraphics[trim={15 3 35 5},clip,width=0.5\textwidth]{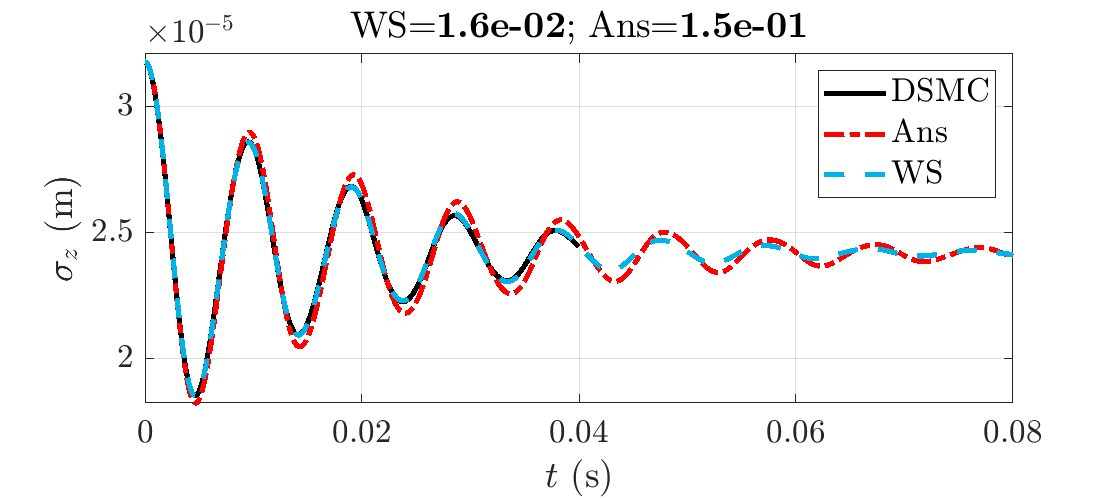}
\end{tabular}

\begin{tabular}{@{}c@{}|c@{}}
\begin{minipage}{0.5\textwidth}
\resizebox{\textwidth}{!}{
\begin{tabular}{p{0.3cm}|p{2cm}|p{0.4cm}|p{1.6cm}|p{1.6cm}|p{1.8cm}}
 & term & AE & Ansatz Coeff & Learned Coeff & term mag \\ \hline
A & $\sigma_\perp$ & 1 & -9.62e+05 & -9.62e+05 & 4.48e-01 \\
& $\sigma_\perp^{-1}$ & 1 & 2.16e-04 & 2.16e-04 & 1.26e+00 \\
& $\sigma_\perp^{-1}\sigma_z^2$ & 1 & -6.16e+04 & -6.16e+04 & 6.29e-01 \\
& $\sigma_\perp^{-1}\dot{\sigma_\perp}^2$ & 1 & -6.67e-01 & -6.67e-01 & 3.44e-04 \\
& $\dot{\sigma}_\perp\sigma_\perp^{-2}$ & 0 & -1.69e-08 & -1.75e-07 & 4.86e-03 \\
& $\sigma_\perp^{-1}\dot{\sigma_z}^2$ & 1 & -3.33e-01 & -3.33e-01 & 9.54e-02 \\
& $\dot{\sigma}_z\sigma_\perp^{-1}\sigma_z^{-1}$ & 0 & 1.69e-08 & 1.68e-08 & 3.15e-02 \\
\hline F & $\sigma_z^2$ & 0 & 0.00e+00 & 1.57e+09 & 1.72e-01 \\
& $\sigma_z^3$ & 0 & 0.00e+00 & -3.56e+13 & 1.27e-01 \\
& $\sigma_\perp^2\sigma_z^{-1}$ & 0 & 0.00e+00 & 4.10e+04 & 6.54e-02 \\
& $\sigma_\perp^2$ & 0 & 0.00e+00 & -3.86e+09 & 6.64e-02 \\
\hline V & $\sigma_\perp^{-2}\dot{\sigma}_\perp\dot{\sigma_\perp}^2$ & 0 & 0.00e+00 & 9.44e-02 & 1.88e-04 \\
& $\dot{\sigma}_\perp\sigma_\perp^{-1}\sigma_z^{-1}$ & 0 & 0.00e+00 & 3.52e-07 & 9.13e-05 \\
& $\sigma_\perp^{-1}\sigma_z^{-1}\dot{\sigma}_\perp\dot{\sigma_\perp}^2$ & 0 & 0.00e+00 & -1.35e-01 & 1.78e-04 \\
& $\dot{\sigma}_\perp$ & 0 & 0.00e+00 & -1.29e+03 & 2.87e-03 \\
& $\sigma_\perp^{-1}\sigma_z^{-1}\dot{\sigma}_\perp\dot{\sigma_z}^2$ & 0 & 0.00e+00 & 1.64e-05 & 4.53e-05 \\
& $\dot{\sigma}_z\sigma_z^{-2}$ & 0 & 0.00e+00 & 2.07e-09 & 2.51e-03 \\
& $\sigma_z^{-2}\dot{\sigma}_\perp\dot{\sigma_z}^2$ & 0 & 0.00e+00 & 5.23e-04 & 8.87e-04 \\
& $\sigma_\perp^{-1}\sigma_z\dot{\sigma}_\perp$ & 0 & 0.00e+00 & 5.92e+02 & 6.23e-03 \\
\end{tabular}}
\end{minipage}&
\begin{minipage}{0.5\textwidth}
\resizebox{\textwidth}{!}{
\begin{tabular}{p{0.3cm}|p{2cm}|p{0.4cm}|p{1.6cm}|p{1.6cm}|p{1.8cm}}
  & term & AE & Ansatz Coeff & Learned Coeff & term mag \\ \hline
A & $\sigma_z$ & 1 & -2.46e+05 & -2.46e+05 & 4.44e-01 \\
& $\sigma_z^{-1}$ & 1 & 2.16e-04 & 2.16e-04 & 9.93e-01 \\
& $\sigma_\perp^2\sigma_z^{-1}$ & 1 & -3.85e+05 & -3.85e+05 & 4.68e-01 \\
& $\sigma_z^{-1}\dot{\sigma_\perp}^2$ & 1 & -6.67e-01 & -6.67e-01 & 3.05e-03 \\
& $\dot{\sigma}_\perp\sigma_\perp^{-1}\sigma_z^{-1}$ & 0 & 3.38e-08 & -1.02e-08 & 3.13e-03 \\
& $\sigma_z^{-1}\dot{\sigma_z}^2$ & 1 & -3.33e-01 & -3.33e-01 & 4.46e-03 \\
& $\dot{\sigma}_z\sigma_z^{-2}$ & 0 & -3.38e-08 & -2.20e-07 & 4.37e-02 \\
\hline F & $\sigma_\perp^{-2}\sigma_z$ & 0 & 0.00e+00 & -2.97e-05 & 4.61e-01 \\
& $\sigma_\perp^{-1}\sigma_z^2$ & 0 & 0.00e+00 & 1.55e+05 & 1.26e+00 \\
& $\sigma_z^{-3}$ & 0 & 0.00e+00 & 2.82e-15 & 6.44e-02 \\
& $\sigma_z^3$ & 0 & -0.00e+00 & -1.98e+14 & 7.31e-01 \\
& $\sigma_\perp\sigma_z^2$ & 0 & 0.00e+00 & -1.96e+13 & 2.24e-02 \\
\hline V & $\dot{\sigma}_\perp\sigma_\perp^{-2}$ & 0 & 0.00e+00 & 2.46e-08 & 1.07e-02 \\
& $\sigma_\perp^{-2}\dot{\sigma}_z\dot{\sigma_\perp}^2$ & 0 & 0.00e+00 & 5.54e-03 & 3.88e-03 \\
& $\sigma_\perp^{-1}\sigma_z^{-1}\dot{\sigma}_z\dot{\sigma_\perp}^2$ & 0 & 0.00e+00 & -1.32e-03 & 6.02e-04 \\
& $\sigma_\perp\sigma_z^{-1}\dot{\sigma}_z$ & 0 & 0.00e+00 & 5.81e+02 & 2.27e-02 \\
& $\dot{\sigma}_z\sigma_\perp^{-1}\sigma_z^{-1}$ & 0 & 0.00e+00 & 8.81e-08 & 1.67e-02 \\
& $\sigma_\perp^{-1}\sigma_z^{-1}\dot{\sigma}_z\dot{\sigma_z}^2$ & 0 & 0.00e+00 & 2.11e-03 & 3.12e-03 \\
& $\sigma_z^{-2}\dot{\sigma}_z\dot{\sigma_z}^2$ & 0 & 0.00e+00 & -3.95e-03 & 7.29e-03 \\
& $\dot{\sigma}_z$ & 0 & -0.00e+00 & -3.06e+02 & 4.13e-03 \\
\end{tabular}}
\end{minipage}
\end{tabular}
\caption{Similar results to Table \ref{tab:lambda1} for $\lambda = 0.32$.}
\label{tab:lambda3}
\end{table}

\begin{table}[ht]
\begin{tabular}{@{}c@{}c@{}}
\includegraphics[trim={15 3 35 5},clip,width=0.5\textwidth]{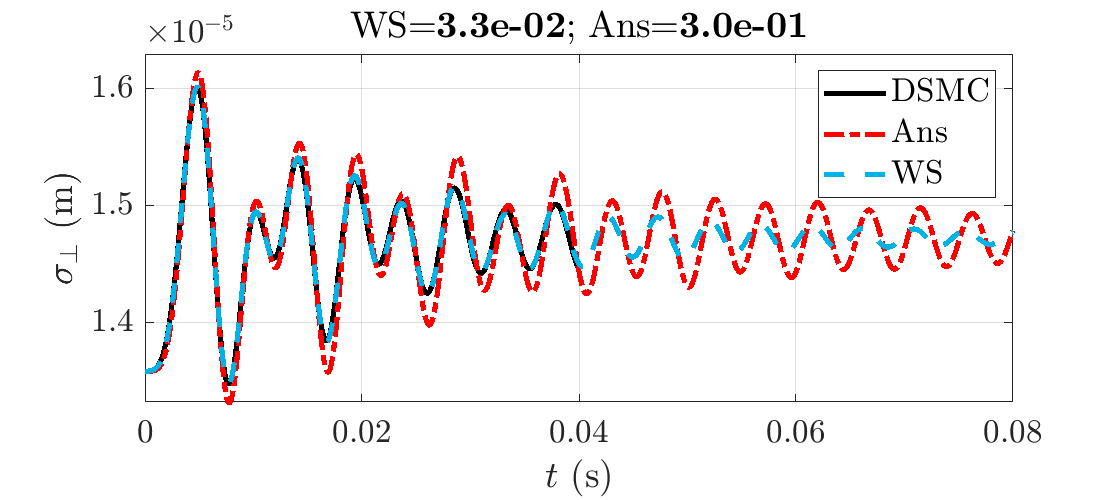} &
\includegraphics[trim={15 3 35 5},clip,width=0.5\textwidth]{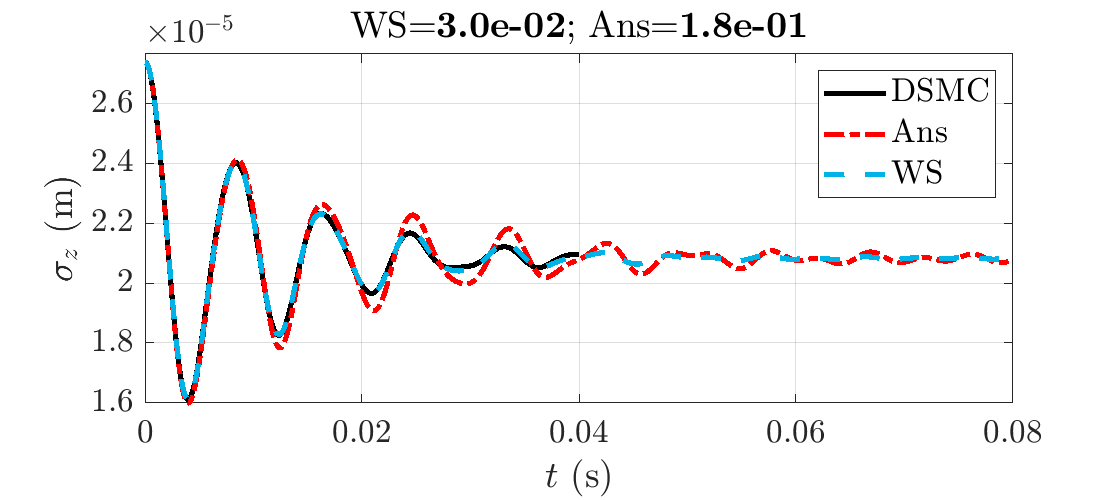}
\end{tabular}

\begin{tabular}{@{}c@{}|c@{}}
\begin{minipage}{0.5\textwidth}
\resizebox{\textwidth}{!}{
\begin{tabular}{p{0.3cm}|p{2cm}|p{0.4cm}|p{1.6cm}|p{1.6cm}|p{1.8cm}}
 & term & AE & Ansatz Coeff & Learned Coeff & term mag \\ \hline
A & $\sigma_\perp$ & 1 & -8.29e+05 & -8.29e+05 & 9.10e-01 \\
& $\sigma_\perp^{-1}$ & 1 & 2.16e-04 & 2.16e-04 & 3.53e-01 \\
& $\sigma_\perp^{-1}\sigma_z^2$ & 1 & -8.29e+04 & -8.29e+04 & 2.62e-01 \\
& $\sigma_\perp^{-1}\dot{\sigma_\perp}^2$ & 1 & -6.67e-01 & -6.67e-01 & 5.48e-04 \\
& $\dot{\sigma}_\perp\sigma_\perp^{-2}$ & 0 & -1.69e-08 & -4.38e-07 & 4.19e-04 \\
& $\sigma_\perp^{-1}\dot{\sigma_z}^2$ & 1 & -3.33e-01 & -3.33e-01 & 5.52e-03 \\
& $\dot{\sigma}_z\sigma_\perp^{-1}\sigma_z^{-1}$ & 0 & 1.69e-08 & -2.86e-09 & 4.65e-03 \\
\hline F & $\sigma_\perp^{-1}\sigma_z^{-1}$ & 0 & 0.00e+00 & 2.36e-10 & 1.41e-02 \\
& $\sigma_z^{-3}$ & 0 & 0.00e+00 & -3.34e-15 & 6.13e-02 \\
& $\sigma_z^2$ & 0 & 0.00e+00 & -1.81e+09 & 4.41e-02 \\
& $\sigma_z^3$ & 0 & 0.00e+00 & 4.38e+13 & 3.88e-02 \\
\hline V & $\sigma_\perp^{-2}\dot{\sigma}_\perp\dot{\sigma_\perp}^2$ & 0 & 0.00e+00 & 1.45e-02 & 5.60e-04 \\
& $\dot{\sigma}_\perp\sigma_\perp^{-1}\sigma_z^{-1}$ & 0 & 0.00e+00 & 6.41e-07 & 9.06e-03 \\
& $\sigma_\perp^{-1}\sigma_z^{-1}\dot{\sigma}_\perp\dot{\sigma_\perp}^2$ & 0 & 0.00e+00 & -2.03e-02 & 1.79e-05 \\
& $\dot{\sigma}_\perp$ & 0 & -0.00e+00 & -2.10e+03 & 8.34e-03 \\
& $\sigma_\perp^{-1}\sigma_z^{-1}\dot{\sigma}_\perp\dot{\sigma_z}^2$ & 0 & 0.00e+00 & 2.40e-03 & 6.55e-03 \\
& $\dot{\sigma}_z\sigma_z^{-2}$ & 0 & 0.00e+00 & 3.05e-08 & 4.27e-02 \\
& $\sigma_z^{-2}\dot{\sigma}_\perp\dot{\sigma_z}^2$ & 0 & 0.00e+00 & -1.08e-03 & 2.32e-03 \\
& $\sigma_\perp^{-1}\sigma_z\dot{\sigma}_\perp$ & 0 & -0.00e+00 & 1.35e+03 & 3.58e-03 \\
\end{tabular}}
\end{minipage}&
\begin{minipage}{0.5\textwidth}
\resizebox{\textwidth}{!}{
\begin{tabular}{p{0.3cm}|p{2cm}|p{0.4cm}|p{1.6cm}|p{1.6cm}|p{1.8cm}}
  & term & AE & Ansatz Coeff & Learned Coeff & term mag \\ \hline
A & $\sigma_z$ & 1 & -3.32e+05 & -3.32e+05 & 3.27e-01 \\
& $\sigma_z^{-1}$ & 1 & 2.16e-04 & 2.16e-04 & 1.12e+00 \\
& $\sigma_\perp^2\sigma_z^{-1}$ & 1 & -3.32e+05 & -3.32e+05 & 4.89e-01 \\
& $\sigma_z^{-1}\dot{\sigma_\perp}^2$ & 1 & -6.67e-01 & -6.67e-01 & 2.11e-03 \\
& $\dot{\sigma}_\perp\sigma_\perp^{-1}\sigma_z^{-1}$ & 0 & 3.38e-08 & 3.77e-08 & 2.15e-02 \\
& $\sigma_z^{-1}\dot{\sigma_z}^2$ & 1 & -3.33e-01 & -3.33e-01 & 4.62e-03 \\
& $\dot{\sigma}_z\sigma_z^{-2}$ & 0 & -3.38e-08 & 1.89e-07 & 5.55e-02 \\
\hline F & $\sigma_\perp^{-1}\sigma_z^{-2}$ & 0 & 0.00e+00 & 4.08e-14 & 1.13e+00 \\
& $\sigma_z^{-3}$ & 0 & 0.00e+00 & 1.05e-14 & 3.36e-01 \\
& $\sigma_z^{-2}$ & 0 & 0.00e+00 & -4.00e-09 & 1.79e+00 \\
& $\sigma_z^3$ & 0 & 0.00e+00 & 7.05e+13 & 1.27e-01 \\
& $\sigma_\perp$ & 0 & 0.00e+00 & -1.04e+04 & 7.00e-03 \\
& $\sigma_\perp^2\sigma_z$ & 0 & 0.00e+00 & -9.16e+14 & 5.31e-02 \\
& $\sigma_\perp^3$ & 0 & 0.00e+00 & 1.67e+15 & 4.17e-01 \\
\hline V & $\dot{\sigma}_\perp\sigma_\perp^{-2}$ & 0 & 0.00e+00 & 3.36e-09 & 2.15e-03 \\
& $\sigma_\perp^{-2}\dot{\sigma}_z\dot{\sigma_\perp}^2$ & 0 & 0.00e+00 & 1.48e-02 & 1.76e-02 \\
& $\sigma_\perp^{-1}\sigma_z^{-1}\dot{\sigma}_z\dot{\sigma_\perp}^2$ & 0 & 0.00e+00 & -1.48e-02 & 1.39e-02 \\
& $\sigma_\perp\sigma_z^{-1}\dot{\sigma}_z$ & 0 & -0.00e+00 & -9.14e+02 & 4.73e-02 \\
& $\dot{\sigma}_z\sigma_\perp^{-1}\sigma_z^{-1}$ & 0 & 0.00e+00 & -1.54e-07 & 3.67e-02 \\
& $\sigma_\perp^{-1}\sigma_z^{-1}\dot{\sigma}_z\dot{\sigma_z}^2$ & 0 & 0.00e+00 & -8.00e-04 & 1.40e-03 \\
& $\sigma_z^{-2}\dot{\sigma}_z\dot{\sigma_z}^2$ & 0 & 0.00e+00 & 1.88e-03 & 4.93e-03 \\
& $\dot{\sigma}_z$ & 0 & 0.00e+00 & 5.77e+02 & 9.83e-03 \\
\end{tabular}}
\end{minipage}
\end{tabular}
\caption{Similar results to Table \ref{tab:lambda1} for $\lambda = 0.5$.}
\label{tab:lambda4}
\end{table}

\begin{table}[ht]
\begin{tabular}{@{}c@{}c@{}}
\includegraphics[trim={15 3 35 5},clip,width=0.5\textwidth]{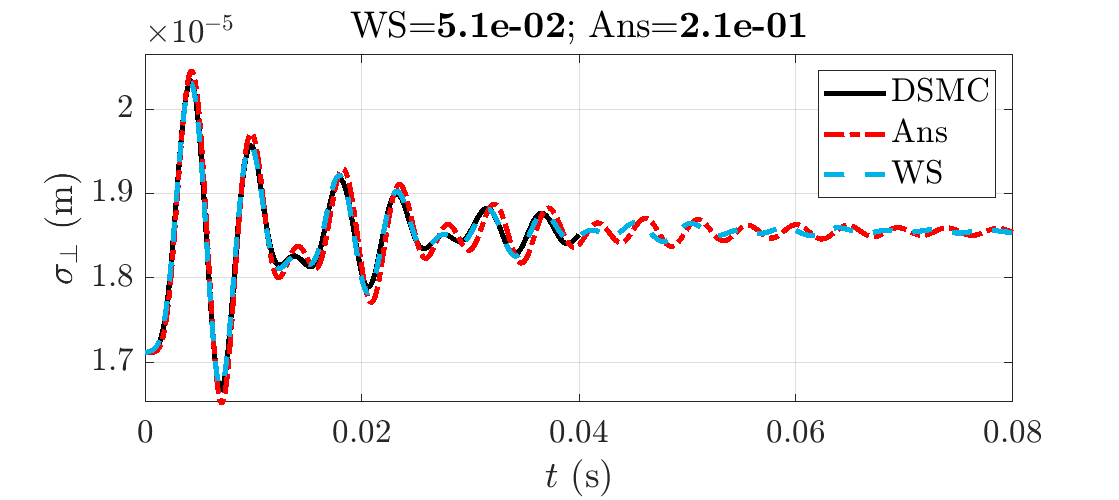} &
\includegraphics[trim={15 3 35 5},clip,width=0.5\textwidth]{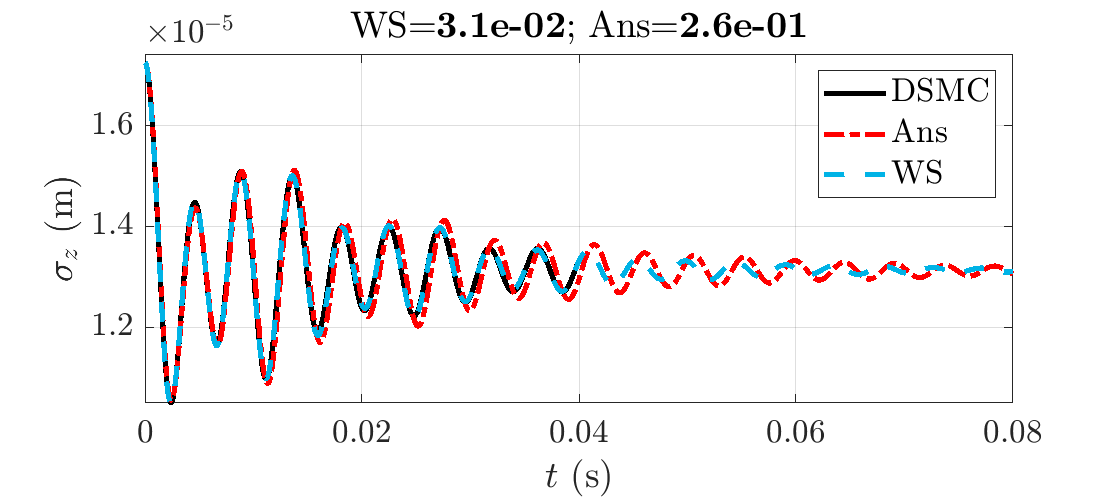}
\end{tabular}

\begin{tabular}{@{}c@{}|c@{}}
\begin{minipage}{0.5\textwidth}
\resizebox{\textwidth}{!}{
\begin{tabular}{p{0.3cm}|p{2cm}|p{0.4cm}|p{1.6cm}|p{1.6cm}|p{1.8cm}}
 & term & AE & Ansatz Coeff & Learned Coeff & term mag \\ \hline
A & $\sigma_\perp$ & 1 & -5.22e+05 & -5.22e+05 & 3.59e-01 \\
& $\sigma_\perp^{-1}$ & 1 & 2.16e-04 & 2.16e-04 & 5.49e-01 \\
& $\sigma_\perp^{-1}\sigma_z^2$ & 1 & -2.09e+05 & -2.09e+05 & 1.17e-01 \\
& $\sigma_\perp^{-1}\dot{\sigma_\perp}^2$ & 1 & -6.67e-01 & -6.67e-01 & 1.12e-03 \\
& $\dot{\sigma}_\perp\sigma_\perp^{-2}$ & 0 & -1.69e-08 & -5.50e-08 & 4.09e-05 \\
& $\sigma_\perp^{-1}\dot{\sigma_z}^2$ & 1 & -3.33e-01 & -3.33e-01 & 8.28e-03 \\
& $\dot{\sigma}_z\sigma_\perp^{-1}\sigma_z^{-1}$ & 0 & 1.69e-08 & 2.18e-08 & 1.28e-02 \\
\hline F & $\sigma_z^{-3}$ & 0 & 0.00e+00 & -1.59e-16 & 1.28e-02 \\
& $\sigma_z^3$ & 0 & 0.00e+00 & 2.28e+13 & 7.98e-03 \\
& $\sigma_\perp^2\sigma_z$ & 0 & 0.00e+00 & 6.24e+12 & 3.73e-03 \\
\hline V & $\sigma_\perp^{-2}\dot{\sigma}_\perp\dot{\sigma_\perp}^2$ & 0 & 0.00e+00 & -3.96e-02 & 2.06e-03 \\
& $\dot{\sigma}_\perp\sigma_\perp^{-1}\sigma_z^{-1}$ & 0 & 0.00e+00 & 2.46e-08 & 1.01e-04 \\
& $\sigma_\perp^{-1}\sigma_z^{-1}\dot{\sigma}_\perp\dot{\sigma_\perp}^2$ & 0 & 0.00e+00 & 2.67e-02 & 2.13e-03 \\
& $\dot{\sigma}_\perp$ & 0 & 0.00e+00 & -3.29e+02 & 1.18e-04 \\
& $\sigma_\perp^{-1}\sigma_z^{-1}\dot{\sigma}_\perp\dot{\sigma_z}^2$ & 0 & -0.00e+00 & -2.84e-03 & 4.18e-03 \\
& $\dot{\sigma}_z\sigma_z^{-2}$ & 0 & 0.00e+00 & -5.86e-09 & 4.40e-03 \\
& $\sigma_z^{-2}\dot{\sigma}_\perp\dot{\sigma_z}^2$ & 0 & 0.00e+00 & 2.39e-03 & 4.31e-03 \\
& $\sigma_\perp^{-1}\sigma_z\dot{\sigma}_\perp$ & 0 & 0.00e+00 & 4.62e+02 & 4.15e-04 \\
\end{tabular}}
\end{minipage}&
\begin{minipage}{0.5\textwidth}
\resizebox{\textwidth}{!}{
\begin{tabular}{p{0.3cm}|p{2cm}|p{0.4cm}|p{1.6cm}|p{1.6cm}|p{1.8cm}}
  & term & AE & Ansatz Coeff & Learned Coeff & term mag \\ \hline
A & $\sigma_z$ & 1 & -8.36e+05 & -8.36e+05 & 2.46e-01 \\
& $\sigma_z^{-1}$ & 1 & 2.16e-04 & 2.16e-04 & 9.41e-01 \\
& $\sigma_\perp^2\sigma_z^{-1}$ & 1 & -2.09e+05 & -2.09e+050& 2.29e-01 \\
& $\sigma_z^{-1}\dot{\sigma_\perp}^2$ & 1 & -6.67e-01 & -6.67e-01 & 4.73e-03 \\
& $\dot{\sigma}_\perp\sigma_\perp^{-1}\sigma_z^{-1}$ & 0 & 3.38e-08 0 -8.70e-11 & 5.00e-06 \\
& $\sigma_z^{-1}\dot{\sigma_z}^2$ & 1 & -3.33e-01 & -3.33e-01 & 2.96e-03 \\
& $\dot{\sigma}_z\sigma_z^{-2}$ & 0 & -3.38e-08 & -4.21e-08 & 9.22e-03 \\
\hline F & $\sigma_\perp^{-1}\sigma_z^2$ & 0 & 0.00e+00 & 3.08e+04 & 1.59e-02 \\
& $\sigma_z^{-3}$ & 0 & 0.00e+00 & 3.43e-16 & 2.20e-02 \\
& $\sigma_z^3$ & 0 & 0.00e+00 & -1.81e+14 & 4.13e-02 \\
\hline V & $\dot{\sigma}_\perp\sigma_\perp^{-2}$ & 0 & 0.00e+00 & 4.25e-08 & 3.07e-03 \\
& $\sigma_\perp^{-2}\dot{\sigma}_z\dot{\sigma_\perp}^2$ & 0 & 0.00e+00 & 1.38e-02 & 2.52e-03 \\
& $\sigma_\perp^{-1}\sigma_z^{-1}\dot{\sigma}_z\dot{\sigma_\perp}^2$ & 0 & 0.00e+00 & -7.95e-03 & 1.74e-03 \\
& $\sigma_\perp\sigma_z^{-1}\dot{\sigma}_z$ & 0 & 0.00e+00 & 1.34e+02 & 4.76e-03 \\
& $\dot{\sigma}_z\sigma_\perp^{-1}\sigma_z^{-1}$ & 0 & 0.00e+00 & 1.15e-08 & 1.11e-03 \\
& $\sigma_\perp^{-1}\sigma_z^{-1}\dot{\sigma}_z\dot{\sigma_z}^2$ & 0 & 0.00e+00 & 6.75e-04 & 4.76e-04 \\
& $\sigma_z^{-2}\dot{\sigma}_z\dot{\sigma_z}^2$ & 0 & 0.00e+00 & -3.14e-04 & 5.78e-04 \\
& $\dot{\sigma}_z$ & 0 & 0.00e+00 & -1.91e+02 & 1.86e-03 \\
\end{tabular}}
\end{minipage}
\end{tabular}
\caption{Similar results to Table \ref{tab:lambda1} for $\lambda = 2$.}
\label{tab:lambda7}
\end{table}

\begin{table}[ht]
\begin{tabular}{@{}c@{}c@{}}
\includegraphics[trim={15 3 35 5},clip,width=0.5\textwidth]{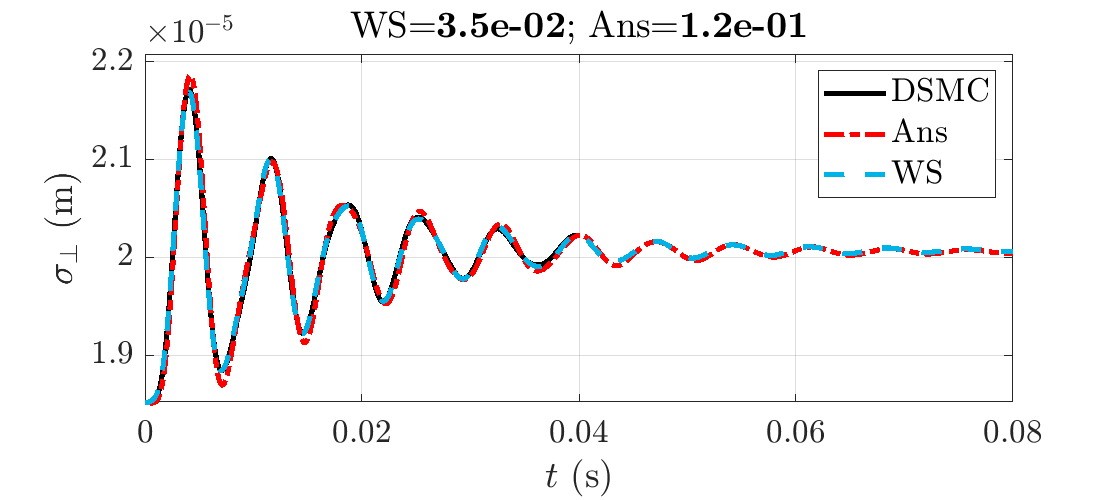} &
\includegraphics[trim={15 3 35 5},clip,width=0.5\textwidth]{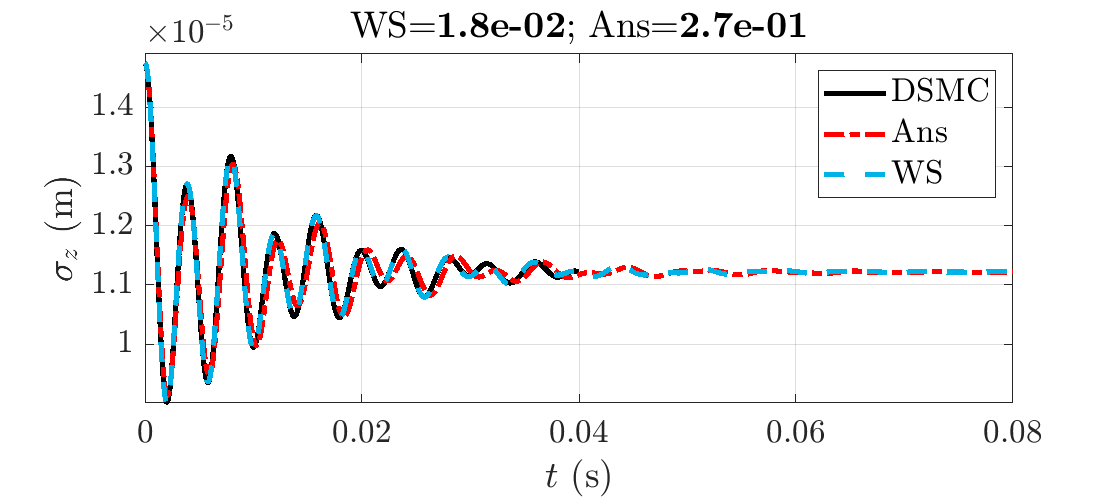}
\end{tabular}

\begin{tabular}{@{}c@{}|c@{}}
\begin{minipage}{0.5\textwidth}
\resizebox{\textwidth}{!}{
\begin{tabular}{p{0.3cm}|p{2cm}|p{0.4cm}|p{1.6cm}|p{1.6cm}|p{1.8cm}}
 & term & AE & Ansatz Coeff & Learned Coeff & term mag \\ \hline
A & $\sigma_\perp$ & 1 & -4.47e+05 & -4.47e+05 & 4.26e-01 \\
& $\sigma_\perp^{-1}$ & 1 & 2.16e-04 & 2.16e-04 & 5.40e-01 \\
& $\sigma_\perp^{-1}\sigma_z^2$ & 1 & -2.86e+05 & -2.86e+05 & 6.74e-02 \\
& $\sigma_\perp^{-1}\dot{\sigma_\perp}^2$ & 1 & -6.67e-01 & -6.67e-01 & 6.95e-04 \\
& $\dot{\sigma}_\perp\sigma_\perp^{-2}$ & 0 & -1.69e-08 & -5.35e-07 & 4.96e-03 \\
& $\sigma_\perp^{-1}\dot{\sigma_z}^2$ & 1 & -3.33e-01 & -3.33e-01 & 1.64e-02 \\
& $\dot{\sigma}_z\sigma_\perp^{-1}\sigma_z^{-1}$ & 0 & 1.69e-08 & 1.26e-08 & 1.08e-02 \\
\hline F & $\sigma_\perp^{-1}\sigma_z^{-2}$ & 0 & 0.00e+00 & -1.59e-16 & 9.90e-03 \\
& $\sigma_\perp\sigma_z^{-2}$ & 0 & 0.00e+00 & -1.87e-07 & 1.80e-03 \\
& $\sigma_\perp\sigma_z^2$ & 0 & 0.00e+00 & 4.15e+13 & 1.46e-02 \\
\hline V & $\sigma_\perp^{-2}\dot{\sigma}_\perp\dot{\sigma_\perp}^2$ & 0 & 0.00e+00 & 3.83e-02 & 5.96e-04 \\
& $\dot{\sigma}_\perp\sigma_\perp^{-1}\sigma_z^{-1}$ & 0 & -0.00e+00 & 2.64e-07 & 1.23e-02 \\
& $\sigma_\perp^{-1}\sigma_z^{-1}\dot{\sigma}_\perp\dot{\sigma_\perp}^2$ & 0 & 0.00e+00 & -2.07e-02 & 2.16e-03 \\
& $\dot{\sigma}_\perp$ & 0 & 0.00e+00 & -1.04e+03 & 2.64e-05 \\
& $\sigma_\perp^{-1}\sigma_z^{-1}\dot{\sigma}_\perp\dot{\sigma_z}^2$ & 0 & 0.00e+00 & 3.20e-03 & 4.88e-03 \\
& $\dot{\sigma}_z\sigma_z^{-2}$ & 0 & 0.00e+00 & -2.77e-10 & 4.37e-04 \\
& $\sigma_z^{-2}\dot{\sigma}_\perp\dot{\sigma_z}^2$ & 0 & 0.00e+00 & -1.61e-03 & 4.95e-03 \\
& $\sigma_\perp^{-1}\sigma_z\dot{\sigma}_\perp$ & 0 & 0.00e+00 & 2.02e+03 & 5.70e-03 \\
\end{tabular}}
\end{minipage}&
\begin{minipage}{0.5\textwidth}
\resizebox{\textwidth}{!}{
\begin{tabular}{p{0.3cm}|p{2cm}|p{0.4cm}|p{1.6cm}|p{1.6cm}|p{1.8cm}}
  & term & AE & Ansatz Coeff & Learned Coeff & term mag \\ \hline
A & $\sigma_z$ & 1 & -1.14e+06 & -1.14e+06 & 2.54e-01 \\
& $\sigma_z^{-1}$ & 1 & 2.16e-04 & 2.16e-04 & 9.65e-01 \\
& $\sigma_\perp^2\sigma_z^{-1}$ & 1 & -1.79e+05 & -1.79e+05 & 2.69e-01 \\
& $\sigma_z^{-1}\dot{\sigma_\perp}^2$ & 1 & -6.67e-01 & -6.67e-01 & 9.35e-03 \\
& $\dot{\sigma}_\perp\sigma_\perp^{-1}\sigma_z^{-1}$ & 0 & 3.38e-08 & 9.53e-09 & 3.18e-04 \\
& $\sigma_z^{-1}\dot{\sigma_z}^2$ & 1 & -3.33e-01 & -3.33e-01 & 3.70e-03 \\
& $\dot{\sigma}_z\sigma_z^{-2}$ & 0 & -3.38e-08 & -4.09e-08 & 1.28e-02 \\
\hline F & $\sigma_z^{-3}$ & 0 & 0.00e+00 & 3.48e-16 & 3.16e-02 \\
& $\sigma_\perp\sigma_z^{-1}$ & 0 & 0.00e+00 & 1.31e-01 & 1.08e-02 \\
& $\sigma_\perp^2\sigma_z$ & 0 & 0.00e+00 & -1.02e+14 & 1.18e-02 \\
\hline V & $\dot{\sigma}_\perp\sigma_\perp^{-2}$ & 0 & 0.00e+00 & 4.02e-08 & 1.48e-03 \\
& $\sigma_\perp^{-2}\dot{\sigma}_z\dot{\sigma_\perp}^2$ & 0 & 0.00e+00 & -4.42e-02 & 1.97e-03 \\
& $\sigma_\perp^{-1}\sigma_z^{-1}\dot{\sigma}_z\dot{\sigma_\perp}^2$ & 0 & 0.00e+00 & 2.58e-02 & 1.78e-03 \\
& $\sigma_\perp\sigma_z^{-1}\dot{\sigma}_z$ & 0 & 0.00e+00 & 2.35e+01 & 1.16e-03 \\
& $\dot{\sigma}_z\sigma_\perp^{-1}\sigma_z^{-1}$ & 0 & 0.00e+00 & 2.46e-08 & 3.01e-03 \\
& $\sigma_\perp^{-1}\sigma_z^{-1}\dot{\sigma}_z\dot{\sigma_z}^2$ & 0 & 0.00e+00 & 2.19e-04 & 1.81e-04 \\
& $\sigma_z^{-2}\dot{\sigma}_z\dot{\sigma_z}^2$ & 0 & 0.00e+00 & 7.71e-05 & 1.93e-04 \\
& $\dot{\sigma}_z$ & 0 & 0.00e+00 & -7.58e+01 & 1.11e-03 \\
\end{tabular}}
\end{minipage}
\end{tabular}
\caption{Similar results to Table \ref{tab:lambda1} for $\lambda = 3.2$.}
\label{tab:lambda8}
\end{table}

\begin{table}[ht]
\begin{tabular}{@{}c@{}c@{}}
\includegraphics[trim={15 3 35 5},clip,width=0.5\textwidth]{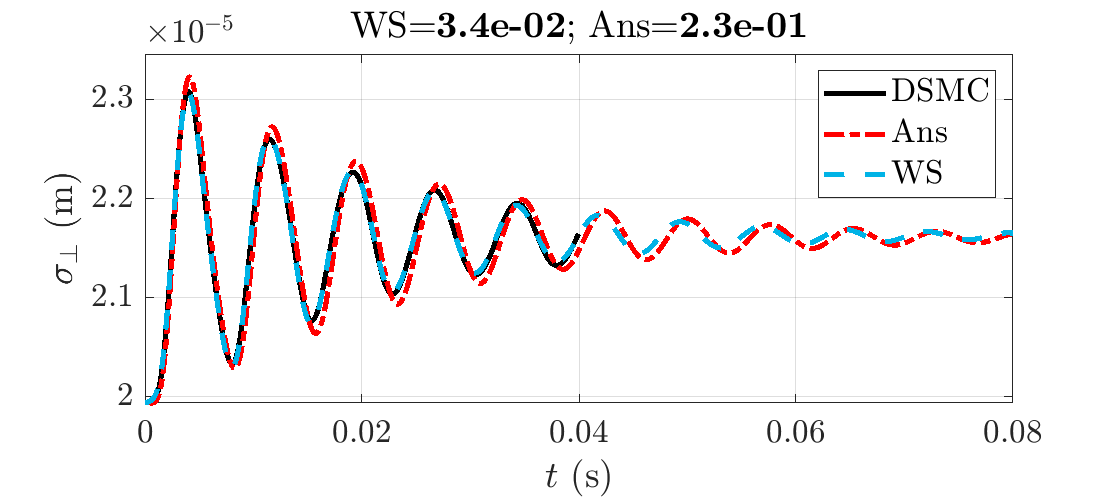} &
\includegraphics[trim={15 3 35 5},clip,width=0.5\textwidth]{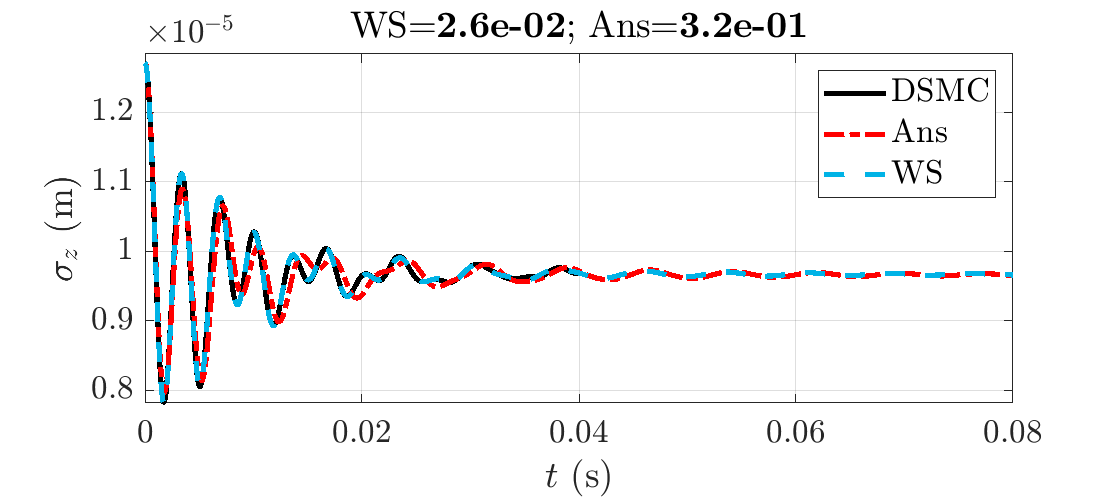}
\end{tabular}

\begin{tabular}{@{}c@{}|c@{}}
\begin{minipage}{0.5\textwidth}
\resizebox{\textwidth}{!}{
\begin{tabular}{p{0.3cm}|p{2cm}|p{0.4cm}|p{1.6cm}|p{1.6cm}|p{1.8cm}}
 & term & AE & Ansatz Coeff & Learned Coeff & term mag \\ \hline
A & $\sigma_\perp$ & 1 & -3.85e+05 & -3.85e+05 & 2.69e-01 \\
& $\sigma_\perp^{-1}$ & 1 & 2.16e-04 & 2.16e-04 & 8.01e-01 \\
& $\sigma_\perp^{-1}\sigma_z^2$ & 1 & -3.85e+05 & -3.85e+05 & 5.67e-02 \\
& $\sigma_\perp^{-1}\dot{\sigma_\perp}^2$ & 1 & -6.67e-01 & -6.67e-01 & 5.89e-04 \\
& $\dot{\sigma}_\perp\sigma_\perp^{-2}$ & 0 & -1.69e-08 & -8.97e-07 & 5.27e-02 \\
& $\sigma_\perp^{-1}\dot{\sigma_z}^2$ & 1 & -3.33e-01 & -3.33e-01 & 7.55e-03 \\
& $\dot{\sigma}_z\sigma_\perp^{-1}\sigma_z^{-1}$ & 0 & 1.69e-08 & 1.21e-08 & 5.87e-03 \\
\hline F & $\sigma_z^{-3}$ & 0 & 0.00e+00 & -1.05e-16 & 1.22e-02 \\
& $\sigma_\perp\sigma_z^2$ & 0 & 0.00e+00 & 6.40e+13 & 1.01e-02 \\
\hline V & $\sigma_\perp^{-2}\dot{\sigma}_\perp\dot{\sigma_\perp}^2$ & 0 & 0.00e+00 & 1.09e-01 & 6.91e-03 \\
& $\dot{\sigma}_\perp\sigma_\perp^{-1}\sigma_z^{-1}$ & 0 & 0.00e+00 & 4.30e-07 & 1.41e-01 \\
& $\sigma_\perp^{-1}\sigma_z^{-1}\dot{\sigma}_\perp\dot{\sigma_\perp}^2$ & 0 & 0.00e+00 & -5.49e-02 & 2.00e-02 \\
& $\dot{\sigma}_\perp$ & 0 & 0.00e+00 & -2.00e+03 & 5.75e-03 \\
& $\sigma_\perp^{-1}\sigma_z^{-1}\dot{\sigma}_\perp\dot{\sigma_z}^2$ & 0 & 0.00e+00 & 3.36e-03 & 9.37e-03 \\
& $\dot{\sigma}_z\sigma_z^{-2}$ & 0 & -0.00e+00 & 7.55e-10 & 1.29e-03 \\
& $\sigma_z^{-2}\dot{\sigma}_\perp\dot{\sigma_z}^2$ & 0 & 0.00e+00 & -8.65e-04 & 6.37e-03 \\
& $\sigma_\perp^{-1}\sigma_z\dot{\sigma}_\perp$ & 0 & 0.00e+00 & 4.04e+03 & 6.44e-02 \\
\end{tabular}}
\end{minipage}&
\begin{minipage}{0.5\textwidth}
\resizebox{\textwidth}{!}{
\begin{tabular}{p{0.3cm}|p{2cm}|p{0.4cm}|p{1.6cm}|p{1.6cm}|p{1.8cm}}
  & term & AE & Ansatz Coeff & Learned Coeff & term mag \\ \hline
A & $\sigma_z$ & 1 & -1.54e+06 & -1.54e+06 & 1.84e-01 \\
& $\sigma_z^{-1}$ & 1 & 2.16e-04 & 2.16e-04 & 1.04e+00 \\
& $\sigma_\perp^2\sigma_z^{-1}$ & 1 & -1.54e+05 & -1.54e+05 & 3.13e-01 \\
& $\sigma_z^{-1}\dot{\sigma_\perp}^2$ & 1 & -6.67e-01 & -6.67e-01 & 2.64e-03 \\
& $\dot{\sigma}_\perp\sigma_\perp^{-1}\sigma_z^{-1}$ & 0 & 3.38e-08 & 1.52e-08 & 3.83e-04 \\
& $\sigma_z^{-1}\dot{\sigma_z}^2$ & 1 & -3.33e-01 & -3.33e-01 & 4.31e-03 \\
& $\dot{\sigma}_z\sigma_z^{-2}$ & 0 & -3.38e-08 & 1.65e-07 & 7.54e-02 \\
\hline F & $\sigma_\perp^{-1}\sigma_z^{-2}$ & 0 & 0.00e+00 & -1.90e-14 & 7.70e-01 \\
& $\sigma_z^{-3}$ & 0 & 0.00e+00 & -2.52e-14 & 3.17e+00 \\
& $\sigma_z^{-2}$ & 0 & 0.00e+00 & 5.14e-09 & 4.34e+00 \\
& $\sigma_z^3$ & 0 & 0.00e+00 & 5.45e+14 & 3.42e-02 \\
& $\sigma_\perp\sigma_z^{-2}$ & 0 & -0.00e+00 & 1.36e-04 & 2.39e+00 \\
& $\sigma_\perp\sigma_z^{-1}$ & 0 & 0.00e+00 & -2.57e+01 & 2.55e+00 \\
& $\sigma_\perp^2\sigma_z$ & 0 & 0.00e+00 & 1.72e+15 & 1.23e-01 \\
\hline V & $\dot{\sigma}_\perp\sigma_\perp^{-2}$ & 0 & 0.00e+00 & 6.22e-08 & 4.34e-04 \\
& $\sigma_\perp^{-2}\dot{\sigma}_z\dot{\sigma_\perp}^2$ & 0 & 0.00e+00 & 1.99e-01 & 9.28e-03 \\
& $\sigma_\perp^{-1}\sigma_z^{-1}\dot{\sigma}_z\dot{\sigma_\perp}^2$ & 0 & -0.00e+00 & -8.84e-02 & 3.93e-03 \\
& $\sigma_\perp\sigma_z^{-1}\dot{\sigma}_z$ & 0 & 0.00e+00 & -9.54e+02 & 6.47e-02 \\
& $\dot{\sigma}_z\sigma_\perp^{-1}\sigma_z^{-1}$ & 0 & 0.00e+00 & -3.67e-07 & 5.50e-02 \\
& $\sigma_\perp^{-1}\sigma_z^{-1}\dot{\sigma}_z\dot{\sigma_z}^2$ & 0 & 0.00e+00 & -1.27e-03 & 1.23e-03 \\
& $\sigma_z^{-2}\dot{\sigma}_z\dot{\sigma_z}^2$ & 0 & 0.00e+00 & 8.89e-04 & 3.13e-03 \\
& $\dot{\sigma}_z$ & 0 & -0.00e+00 & 1.77e+03 & 3.19e-02 \\
\end{tabular}}
\end{minipage}
\end{tabular}
\caption{Similar results to Table \ref{tab:lambda1} for $\lambda = 5$.}
\label{tab:lambda9}
\end{table}

\begin{table}[ht]
\begin{tabular}{@{}c@{}c@{}}
\includegraphics[trim={15 3 35 5},clip,width=0.5\textwidth]{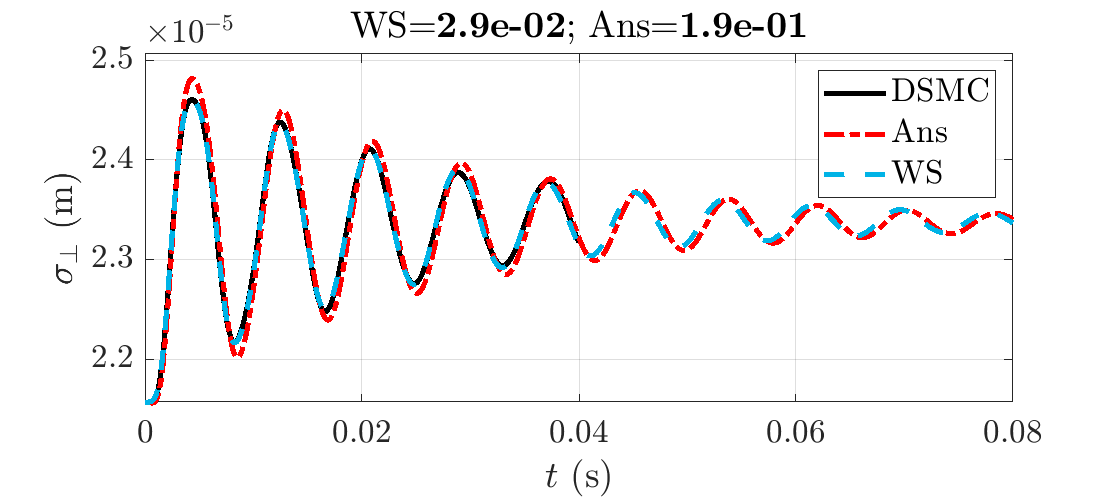} &
\includegraphics[trim={15 3 35 5},clip,width=0.5\textwidth]{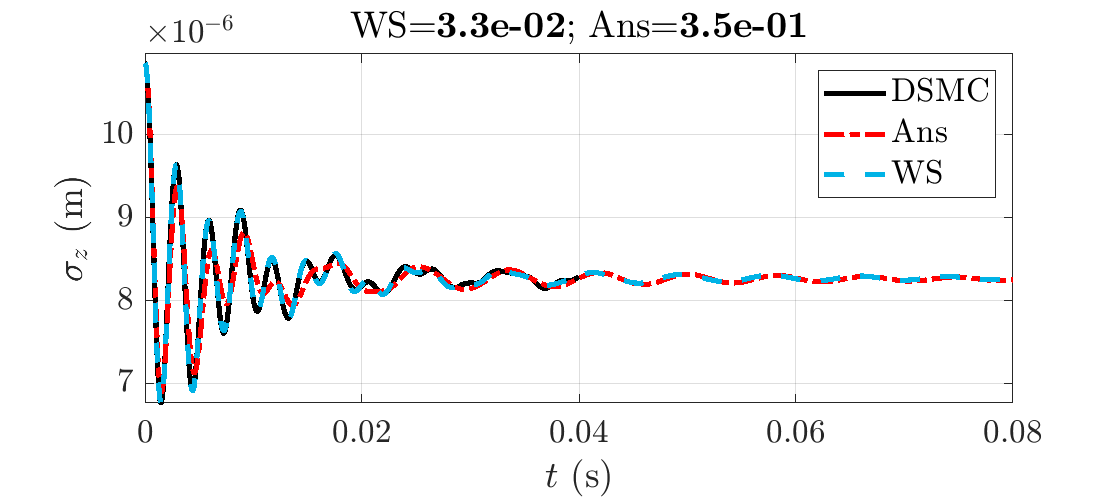}
\end{tabular}

\begin{tabular}{@{}c@{}|c@{}}
\begin{minipage}{0.5\textwidth}
\resizebox{\textwidth}{!}{
\begin{tabular}{p{0.3cm}|p{2cm}|p{0.4cm}|p{1.6cm}|p{1.6cm}|p{1.8cm}}
 & term & AE & Ansatz Coeff & Learned Coeff & term mag \\ \hline
A & $\sigma_\perp$ & 1 & -3.29e+05 & -3.29e+05 & 8.01e-01 \\
& $\sigma_\perp^{-1}$ & 1 & 2.16e-04 & 2.16e-04 & 1.76e-01 \\
& $\sigma_\perp^{-1}\sigma_z^2$ & 1 & -5.26e+05 & -5.26e+05 & 3.92e-02 \\
& $\sigma_\perp^{-1}\dot{\sigma_\perp}^2$ & 1 & -6.67e-01 & -6.67e-01 & 5.25e-04 \\
& $\dot{\sigma}_\perp\sigma_\perp^{-2}$ & 0 & -1.69e-08 & -9.27e-07 & 4.40e-02 \\
& $\sigma_\perp^{-1}\dot{\sigma_z}^2$ & 1 & -3.33e-01 & -3.33e-01 & 1.04e-02 \\
& $\dot{\sigma}_z\sigma_\perp^{-1}\sigma_z^{-1}$ & 0 & 1.69e-08 & 1.35e-08 & 1.12e-02 \\
\hline F & $\sigma_\perp^{-3}$ & 0 & 0.00e+00 & 4.08e-16 & 4.66e-03 \\
& $\sigma_\perp^{-1}\sigma_z^{-2}$ & 0 & 0.00e+00 & 1.42e-16 & 6.35e-03 \\
& $\sigma_z^{-3}$ & 0 & 0.00e+00 & -1.95e-16 & 1.13e-02 \\
& $\sigma_\perp\sigma_z^{-2}$ & 0 & 0.00e+00 & 6.80e-07 & 1.41e-02 \\
\hline V & $\sigma_\perp^{-2}\dot{\sigma}_\perp\dot{\sigma_\perp}^2$ & 0 & 0.00e+00 & -2.88e-02 & 1.15e-03 \\
& $\dot{\sigma}_\perp\sigma_\perp^{-1}\sigma_z^{-1}$ & 0 & 0.00e+00 & 3.22e-07 & 7.78e-02 \\
& $\sigma_\perp^{-1}\sigma_z^{-1}\dot{\sigma}_\perp\dot{\sigma_\perp}^2$ & 0 & 0.00e+00 & 5.91e-03 & 1.57e-03 \\
& $\dot{\sigma}_\perp$ & 0 & -0.00e+00 & -1.86e+03 & 1.12e-02 \\
& $\sigma_\perp^{-1}\sigma_z^{-1}\dot{\sigma}_\perp\dot{\sigma_z}^2$ & 0 & 0.00e+00 & 4.87e-03 & 1.39e-02 \\
& $\dot{\sigma}_z\sigma_z^{-2}$ & 0 & 0.00e+00 & -7.77e-10 & 2.06e-03 \\
& $\sigma_z^{-2}\dot{\sigma}_\perp\dot{\sigma_z}^2$ & 0 & 0.00e+00 & -1.59e-03 & 1.40e-02 \\
& $\sigma_\perp^{-1}\sigma_z\dot{\sigma}_\perp$ & 0 & 0.00e+00 & 5.22e+03 & 2.47e-02 \\
\end{tabular}}
\end{minipage}&
\begin{minipage}{0.5\textwidth}
\resizebox{\textwidth}{!}{
\begin{tabular}{p{0.3cm}|p{2cm}|p{0.4cm}|p{1.6cm}|p{1.6cm}|p{1.8cm}}
  & term & AE & Ansatz Coeff & Learned Coeff & term mag \\ \hline
A & $\sigma_z$ & 1 & -2.11e+06 & -2.11e+06 & 1.54e-01 \\
& $\sigma_z^{-1}$ & 1 & 2.16e-04 & 2.16e-04 & 1.09e+00 \\
& $\sigma_\perp^2\sigma_z^{-1}$ & 1 & -1.32e+05 & -1.32e+05 & 3.31e-01 \\
& $\sigma_z^{-1}\dot{\sigma_\perp}^2$ & 1 & -6.67e-01 & -6.67e-01 & 8.71e-04 \\
& $\dot{\sigma}_\perp\sigma_\perp^{-1}\sigma_z^{-1}$ & 0 & 3.38e-08 & 1.41e-07 & 3.04e-03 \\
& $\sigma_z^{-1}\dot{\sigma_z}^2$ & 1 & -3.33e-01 & -3.33e-01 & 4.43e-03 \\
& $\dot{\sigma}_z\sigma_z^{-2}$ & 0 & -3.38e-08 & -7.95e-08 & 4.42e-02 \\
\hline F & $\sigma_\perp^{-1}\sigma_z^{-2}$ & 0 & 0.00e+00 & -1.84e-15 & 8.24e-02 \\
& $\sigma_z^{-3}$ & 0 & 0.00e+00 & 5.54e-16 & 9.61e-02 \\
& $\sigma_z^3$ & 0 & 0.00e+00 & 6.71e+14 & 2.22e-02 \\
& $\sigma_\perp\sigma_z^{-2}$ & 0 & 0.00e+00 & 2.78e-06 & 6.33e-02 \\
& $\sigma_\perp^2\sigma_z$ & 0 & -0.00e+00 & -2.48e+14 & 1.25e-02 \\
\hline V & $\dot{\sigma}_\perp\sigma_\perp^{-2}$ & 0 & 0.00e+00 & -3.01e-07 & 2.57e-03 \\
& $\sigma_\perp^{-2}\dot{\sigma}_z\dot{\sigma_\perp}^2$ & 0 & 0.00e+00 & -4.71e-02 & 1.33e-03 \\
& $\sigma_\perp^{-1}\sigma_z^{-1}\dot{\sigma}_z\dot{\sigma_\perp}^2$ & 0 & 0.00e+00 & 2.04e-02 & 7.73e-04 \\
& $\sigma_\perp\sigma_z^{-1}\dot{\sigma}_z$ & 0 & 0.00e+00 & 3.50e+02 & 2.77e-02 \\
& $\dot{\sigma}_z\sigma_\perp^{-1}\sigma_z^{-1}$ & 0 & 0.00e+00 & 1.23e-07 & 1.84e-02 \\
& $\sigma_\perp^{-1}\sigma_z^{-1}\dot{\sigma}_z\dot{\sigma_z}^2$ & 0 & 0.00e+00 & 3.51e-03 & 3.34e-03 \\
& $\sigma_z^{-2}\dot{\sigma}_z\dot{\sigma_z}^2$ & 0 & 0.00e+00 & -1.51e-03 & 6.27e-03 \\
& $\dot{\sigma}_z$ & 0 & 0.00e+00 & -7.84e+02 & 1.41e-02 \\
\end{tabular}}
\end{minipage}
\end{tabular}
\caption{Similar results to Table \ref{tab:lambda1} for $\lambda = 8$.}
\label{tab:lambda10}
\end{table}

\end{document}